\documentclass[pdflatex,sn-nature,twocolumn]{sn-jnl}

\usepackage{graphicx}%
\usepackage{multirow}%
\usepackage{amsmath,amssymb,amsfonts}%
\usepackage{amsthm}%
\usepackage{mathrsfs}%
\usepackage[title]{appendix}%
\usepackage{xcolor}%
\usepackage{textcomp}%
\usepackage{manyfoot}%
\usepackage{booktabs}%
\usepackage{algorithm}%
\usepackage{algorithmicx}%
\usepackage{algpseudocode}%
\usepackage{listings}%
\usepackage{physics}
\usepackage{hyperref}
\usepackage{tikz}
\usepackage{graphicx}
\usepackage{svg}
\usepackage{subcaption}
\usepackage{dblfloatfix}
\usepackage{xcolor}
\usepackage[switch]{lineno}

\usepackage{multicol}
\usepackage{geometry}
\theoremstyle{thmstyleone}%
\theoremstyle{thmstyletwo}%

\theoremstyle{thmstylethree}%

\begin{document}

\title{Minimum measurements quantum protocol for band structure calculations}



\author*[1,2]{\fnm{Michal} \sur{Krej{\v c}{\' i}}}

\author[1]{\fnm{Lucie} \sur{Krej{\v c}{\' i}}}

\author[3]{\fnm{Ijaz Ahamed} \sur{Mohammad}}

\author[3,4]{\fnm{Martin} \sur{Plesch}}

\author[1,2]{\fnm{Martin} \sur{Fri{\' a}k}}
\email{friak@ipm.cz, mafri@mail.muni.cz}

\affil[1]{\orgdiv{Institute of Physics of Materials, v. v. i.}, \orgname{Czech Academy of Sciences}, \orgaddress{\street{\v{Z}i\v{z}kova 22}, \city{Brno}, \postcode{616~00}, \country{Czech Republic}}}

\affil[2]{\orgdiv{Department of Condensed Matter Physics, Faculty of Science}, \orgname{Masaryk University}, \orgaddress{\street{ Kotl\'{a}\v{r}sk\'{a} 2}, \city{Brno}, \postcode{611~37}, \country{Czech Republic}}}

\affil[3]{\orgdiv{Institute of Physics}, \orgname{Slovak Academy of Sciences}, \orgaddress{\street{D\'{u}bravsk\'{a} cesta 9}, \city{Bratislava 45}, \postcode{845~11}, \country{Slovak Republic}}}

\affil[4]{\orgdiv{Faculty of Natural Sciences}, \orgname{Matej Bel University}, \orgaddress{\street{ Tajovsk\'{e}ho 40}, \city{Bansk\'{a} Bystrica}, \postcode{974~09}, \country{Slovak Republic}}}


\abstract{Protocols for quantum measurement are an essential part of quantum computing. Measurements are no longer confined to the final step of computation but are increasingly embedded within quantum circuits as integral components of noise-resilient algorithms. However, each observable typically requires a distinct measurement basis, often demanding a different circuit configuration. As the number of such configurations typically grows with the number of qubits, measurements constitute a major bottleneck. Focusing on electronic structure calculations in crystalline systems, we propose a measurement protocol that \textcolor{black}{restricts the required measurement configurations to an absolute minimum of just three}, independent of the number of qubits. This makes it one of the few known protocols that do not scale with qubit number. In particular, we derive the measurement protocol from the symmetries of tight-binding (TB) Hamiltonians and implement it within the \textcolor{black}{Orthogonal-Ansatz Variational Quantum Eigensolver (OA-VQE)} algorithm. We demonstrate its performance on three systems, namely a two-dimensional CuO$_2$ square lattice (3 qubits), bilayer graphene with \textcolor{black}{hexagonal (Honeycomb) lattice} (4 qubits) \textcolor{black}{and three-dimensional diamond lattice (10 qubits)}. \textcolor{black}{Beyond tight-binding systems, the protocol can be extended to enable efficient initial state preparation for many-body Hamiltonians, such as multi-orbital Hubbard models in a momentum space.}}


\maketitle

\section{Introduction}\label{sec1}
\textcolor{black}{At the heart of quantum computing, particularly in the simulation of strongly correlated electronic structures and quantum materials, lies the efficiency of the underlying measurement protocols ~\cite{Ge_2025,Burns_2025,Caprotti_2024,Ballesteros-Ferraz_2024,Romanova_2026,Altuntas_2025,Kaldenbach_2025,Lee_2024,Nie_2024}.} Beyond simply reading out the final answer, strategically placed and high-fidelity measurements are essential resources that {\it enable} computation and reliability. 
In particular, mid-circuit measurements play~\cite{cj89-4h5t} a crucial role in stabilizing logical qubits, i.e., paving a path towards the end of the current Noisy Intermediate-Scale Quantum (NISQ) era.

While the measurement protocols are clearly established as a crucial, non-optional part of any useful quantum computation~\cite{Nature-Phys-2023-Zhu},
the measurement process represents a significant bottleneck -- each observable often requires a distinct measurement basis, necessitating a different setup of the quantum circuit. To address this challenge, we propose a measurement protocol that maximally reduces the number of measurement settings to exactly three, i.e., constant overhead. As the number of proposed measurement settings is constant, independently of the number of qubits, the reduction increases effectively with the increasing number of qubits. The proposed reduction is particularly important in scenarios where state preparation is nontrivial~\cite{plesch2011quantum}. In photonic quantum systems, for instance, modifying or reconfiguring the quantum circuit often requires considerably more effort than in other architectures. Moreover, in near-term quantum computing applications, where execution cost is influenced by the number of circuit runs or state preparations, excessive measurement requirements can significantly increase computational expenses for users.  

Our proposed measurement protocol is related to the calculation of the electronic structure of materials. Calculations of the electronic structure remain a central challenge in condensed matter physics and quantum chemistry. Conventional classical approaches, such as density functional theory (DFT), provide powerful tools for investigating material properties, but often face limitations when applied to highly correlated systems or very large problems~\cite{Schuch, Cohen, Whitfield}. The tight-binding (TB) model provides a valuable semi-empirical alternative, capturing essential electronic features with reduced computational overhead, making it an attractive candidate for exploring quantum computation strategies~\cite{SK, Chadi, Harrison}. Hybrid quantum-classical algorithms, most notably the variational quantum eigensolver (VQE)~\cite{Peruzzo} and its extension, the Variational Quantum Deflation (VQD) algorithm~\cite{Higgott}, \textcolor{black}{Subspace-Search Variational Quantum Eigensolver (SSVQE)~\cite{Nakanishi}, or Orthogonal-Ansatz Variational Quantum Eigensolver (OA-VQE)~\cite{Sherbert2}} have emerged as leading candidates for simulating quantum systems on near-term hardware. However, two key bottlenecks remain: the large number of measurement settings required to evaluate observables and the cost of classical optimisation in high-dimensional parameter spaces~\cite{Bittel, Bonet, Kandala, McClean}. For tight-binding Hamiltonians, existing protocols typically require \(\mathcal{O}(N)\) distinct Pauli measurement settings, where \(N\) is the number of qubits~\cite{Sherbert1, Sherbert2}.

In this work, we present an improvement to the \textcolor{black}{VQE} framework tailored to TB Hamiltonians. By exploiting symmetries in the TB model, we rigorously prove that only three distinct measurement bases are sufficient to evaluate all required observables, yielding a constant measurement overhead, independent of system size. For example, in a near-term practical implementation of a 1000-qubit tight-binding (TB) system—which would typically require thousands of measurement settings—our approach reduces this number to just three. \textcolor{black}{Furthermore, this protocol can be extended to construct exact eigenstates of a many-electron non-interacting system for multi-orbital Hubbard models in momentum space, such as the three-band Emery model. These wavefunctions can subsequently serve as high-fidelity initial states for advanced quantum computing routines, including adiabatic state preparation designed to slowly evolve the non-interacting reference into the true ground state of a fully interacting system.}
 
Although the classical simulations of tight-binding models will likely remain more efficient than their quantum counterparts in the near future, the significance of our contribution lies in showing that, unlike classical computational cost, the number of measurement settings in variational quantum algorithms can be made independent of system size. This property highlights a promising direction for quantum algorithms, where measurement bottlenecks can be overcome even in the NISQ era.


\section[Tight-Binding Model and Hamiltonian Formulation]{Tight-Binding Model and Hamiltonian Formulation}\label{sec2}

The tight-binding (TB) model~\cite{SK} is a widely used semi-empirical method in solid state physics to describe the behaviour of electrons in the periodic potential of a crystal lattice. The core idea is that electrons remain largely localised around individual atoms but can tunnel, or `hop,' to neighbouring sites due to the finite overlap of atomic orbitals. As a single-electron approximation, the TB model captures the essential features of electronic band formation while maintaining relatively low computational complexity. Given its well-established role in the field~\cite{SK, Chadi, Goringe}, we present here only the specific equations relevant to our analysis. An important point is that the electron is embedded in a periodic potential created by the ions arranged in the crystal lattice. Owing to the translational symmetry of the crystal lattice, it is convenient to express the TB Hamiltonian in a \textcolor{black}{momentum space}, where it takes the form

\begin{equation}
    \label{eq.h_rec}
    \mathcal{\hat{H}}(\boldsymbol{k}) = \sum_{j} \varepsilon_{j}\hat{c}^{\dagger}_{j}\hat{c}^{\textcolor{white}{\dagger}}_{j} + \sum_{j,l \atop j \neq l}\mathcal{H}_{jl}(\boldsymbol{k})\hat{c}^{\dagger}_{j}\hat{c}^{\textcolor{white}{\dagger}}_{l},
\end{equation}

\noindent where $\boldsymbol{k}$ is the Bloch wave vector, the indices $j,l$ label the atomic orbitals within the crystal unit cell, the operators $\hat{c}^{\dagger}_{j}$ and $\hat{c}^{\textcolor{white}{\dagger}}_{j}$ are the creation and annihilation operators \textcolor{black}{that respectively add or remove an electron in the single-particle orbital $j$}. The first term in the Hamiltonian~\eqref{eq.h_rec} represents the on-site energies of the occupied orbitals. The second term accounts for the hopping processes in which the electron hops from one orbital to another. Here, the matrix elements $\mathcal{H}_{jl}(\boldsymbol{k})$ are given by the following formula.

\begin{equation}
    \label{eq.h_matrix_el}
    \mathcal{H}_{jl}(\boldsymbol{k}) = \sum_{\boldsymbol{R}}t_{jl}\text{e}^{i\boldsymbol{k}\cdot \boldsymbol{R}}.
\end{equation}

\noindent The sum $\sum_{\boldsymbol{R}}$ runs over the first or second nearest neighbours, depending on the specific model. The parameters $t_{jl}$ are the hopping amplitudes that describe the energy associated with an electron hopping between individual orbitals. The important parameters, namely, on-site energies $\varepsilon_{j}$ and hopping amplitudes $t_{jl}$, are model-dependent and are typically obtained from more sophisticated ab initio methods, such as Density Functional Theory (DFT), or from experimental data. In DFT, the electronic problem is often reformulated into a tight-binding-like model by expressing it in terms of localised Wannier orbitals and effective hopping parameters. This extends the usefulness of the tight-binding model, making it applicable to a broader range of materials and systems~\cite{Wannier}.

Anyway, using the \textcolor{black}{momentum} space representation, one only needs to diagonalise the $N\times N$ Hamiltonian matrix, where $N$ denotes the number of orbitals in the unit cell. This diagonalisation is performed separately for all $\boldsymbol{k}$-points that are to be evaluated, in our case along the high-symmetry path defining the usual band structure in the first Brillouin zone of the reciprocal lattice.

To calculate the energy eigenvalues on a quantum computer, one must first select an appropriate mapping between the creation/annihilation operators and the qubit operators. The simplest and most convenient approach is the orbital-qubit mapping introduced in~\cite{Sherbert1, Sherbert2}, where the qubit states $\ket{0}$ and $\ket{1}$ are associated with the occupation of an orbital by an electron. The states $\ket{0}$, $\ket{1}$ represent the unoccupied and occupied orbitals, respectively. \textcolor{black}{Due to the single particle nature of the problem, one does not have to enforce the anticommutativity of qubit operators via chains of Jordan-Wigner (JW) $\hat{Z}$ strings}. In particular, the mapping between the creation/annihilation operators and qubit operators is achieved by \textcolor{black}{the single-particle JW transformations}

\begin{equation}
    \label{eq.mapping}
    \hat{c}^{\dagger}_{j} = \dfrac{1}{2}\left(\hat{X}_j - i\hat{Y}_{j}\right), \quad \hat{c}^{\textcolor{white}{\dagger}}_{j} = \dfrac{1}{2}\left(\hat{X}_{j} + i\hat{Y}_{j}\right),
\end{equation}

The corresponding qubit Hamiltonian can be written as

\begin{align}
\label{eq.ham_qubit}
\hat{\mathcal{H}}(\boldsymbol{k})
& = \dfrac{1}{2} \sum_{j} \varepsilon_j (\hat{I} - \hat{Z}_{j}) \nonumber \\
& + \dfrac{1}{2} \sum_j \sum_{l>j} \text{Re}\left\{\mathcal{H}_{jl}(\boldsymbol{k}) \right\} 
\left( \hat{X}_{j} \hat{X}_{l} + \hat{Y}_{j} \hat{Y}_{l} \right) \nonumber \\
& + \dfrac{1}{2} \sum_j \sum_{l>j} \text{Im}\left\{ \mathcal{H}_{jl}(\boldsymbol{k}) \right\}
\left( \hat{Y}_{j} \hat{X}_{l} - \hat{X}_{j} \hat{Y}_{l} \right),
\end{align}

\begin{figure*}[b!]
  \centering
    \begin{subfigure}{0.47\textwidth}
        \centering
        \includegraphics[width=\linewidth]{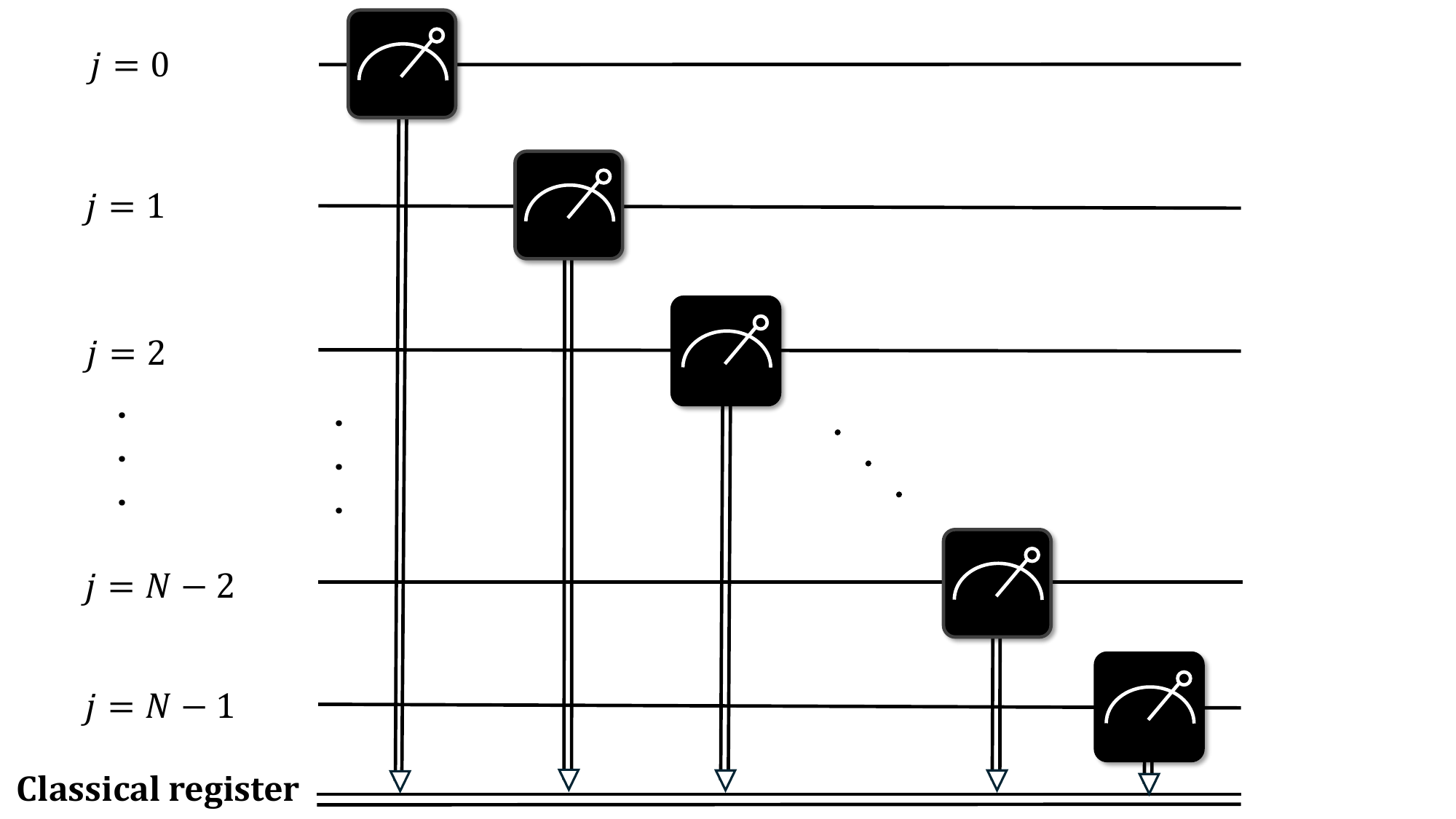}
        \caption{$\mathcal{M}_{Z}$}
        \label{fig:Mz}
    \end{subfigure}
    \hfill
    \begin{subfigure}{0.47\textwidth}
        \centering
        \includegraphics[width=\linewidth]{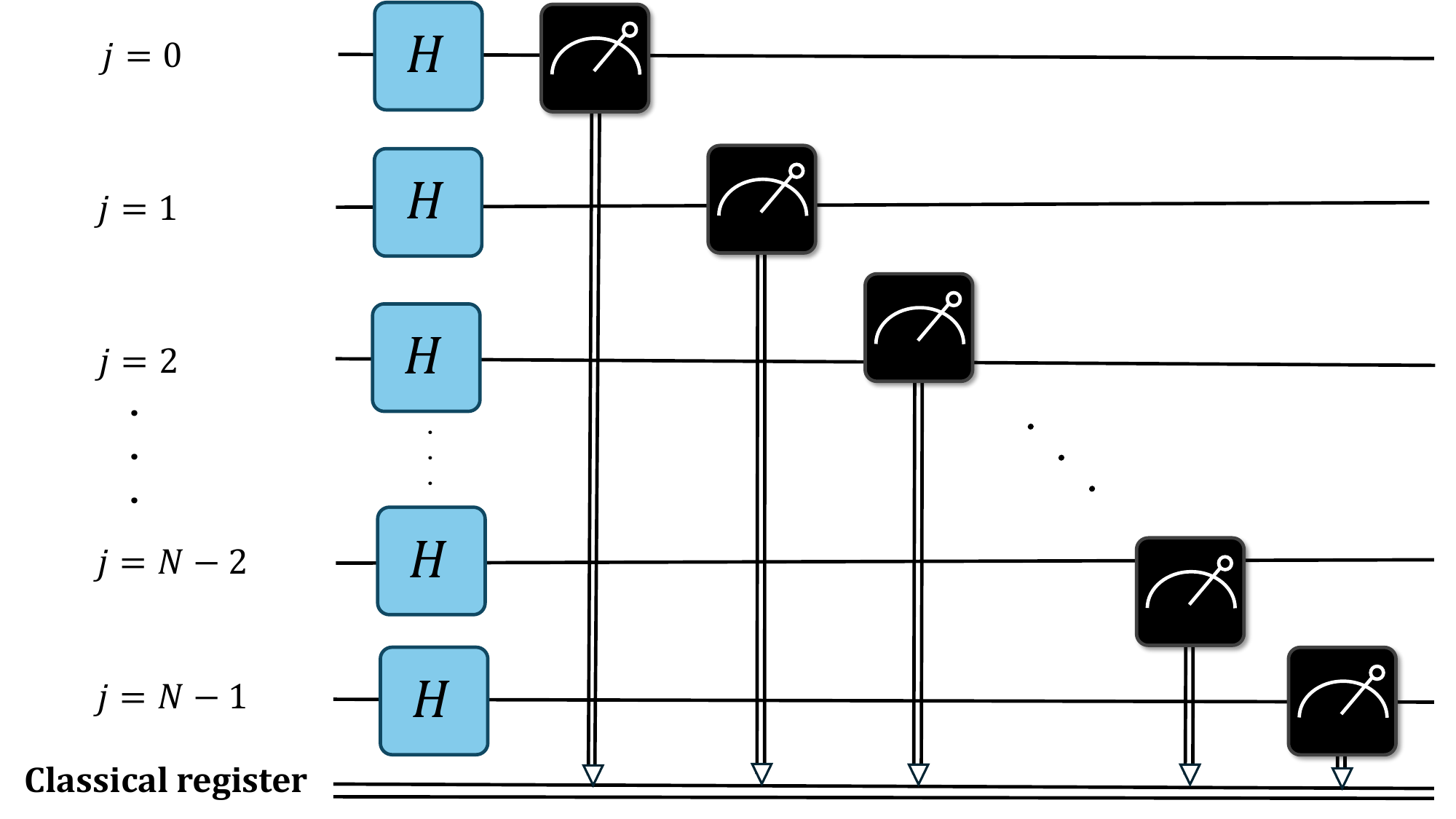}
        \caption{$\mathcal{M}_{XX}$}
        \label{fig:Mxx}
    \end{subfigure}
    
    \vspace{0.3cm} 
    
    \begin{subfigure}{0.47\textwidth}
        \centering
        \includegraphics[width=\linewidth]{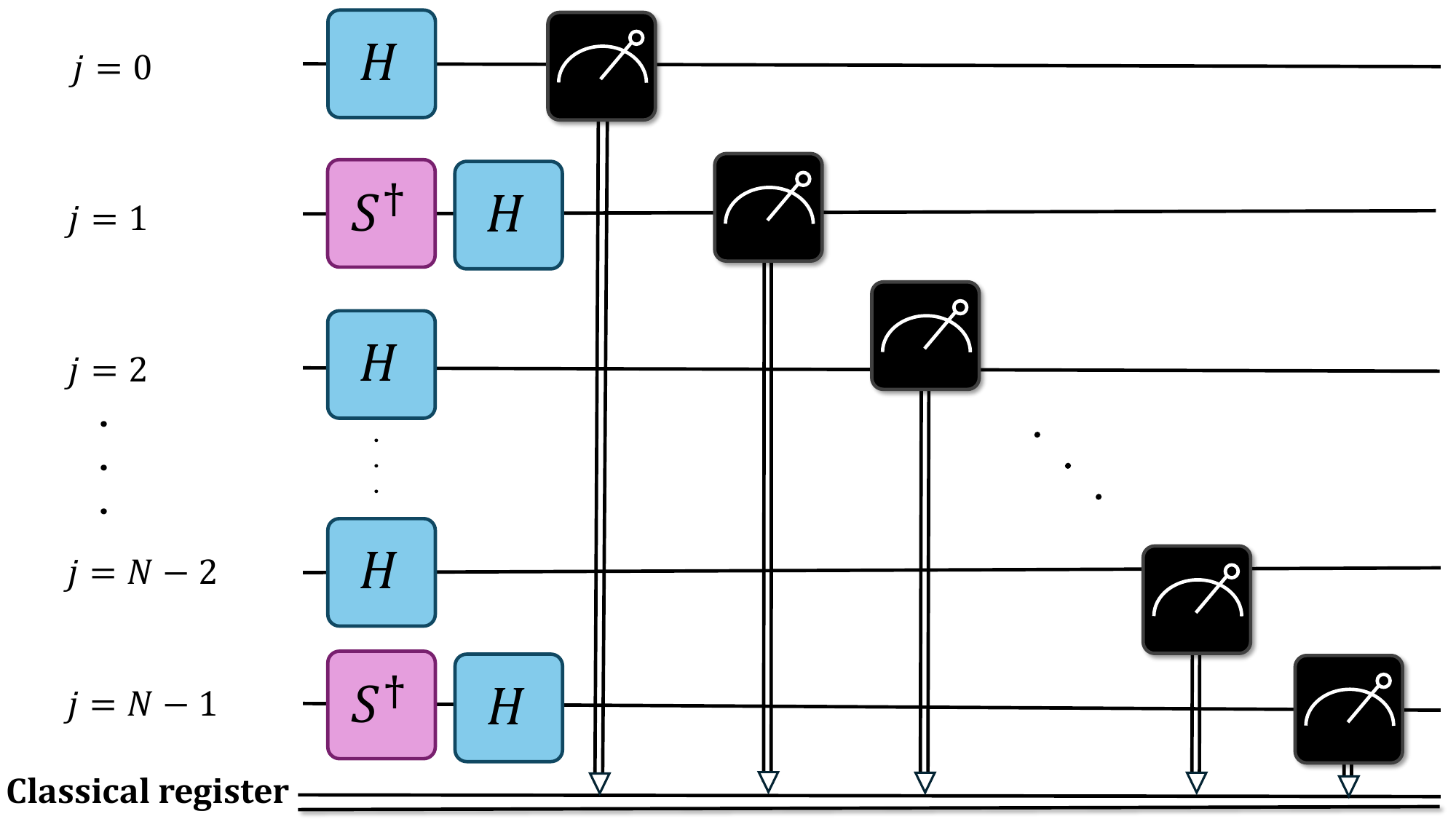}
        \caption{$\mathcal{M}_{XY}$}
        \label{fig:Mxy}
    \end{subfigure}
    \hfill
    \begin{subfigure}{0.47\textwidth}
        \centering
        \includegraphics[width=\linewidth]{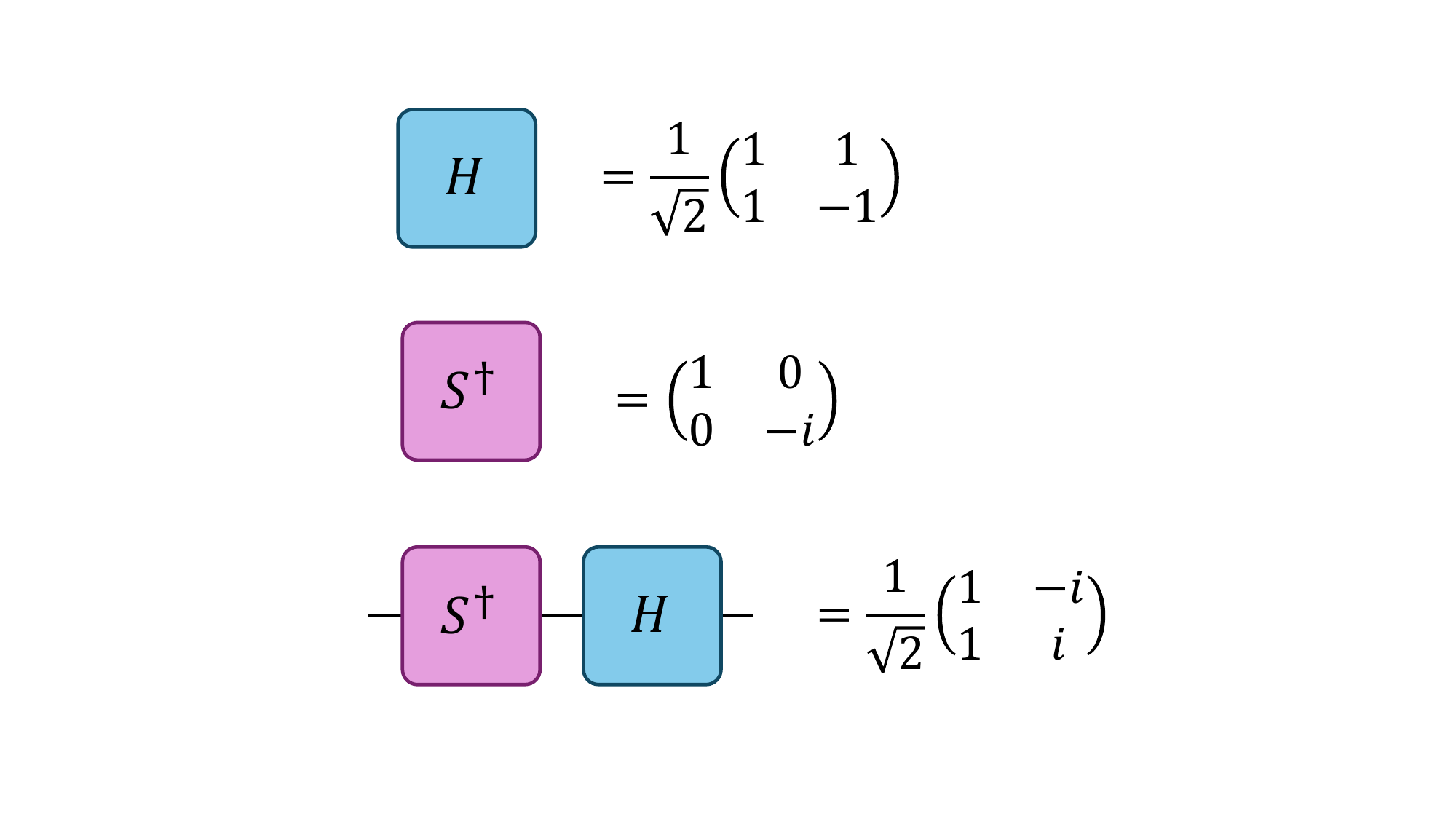}
        \caption{Measurement gates.}
        \label{fig:meas}
    \end{subfigure}

    \caption{Three measurement circuits $\mathcal{M}_{Z}$, $\mathcal{M}_{XX}$, $\mathcal{M}_{XY}$ for cost function, Eq.~\eqref{eq.cost_fun}, estimation. (a) The measurement circuit for estimating the probabilities $\abs{a_{j}}^{2},\, j=0,1,\ldots,N-1$. 
    (b) The measurement circuit for estimating all $\langle\hat{X}_{j}\hat{X}_{l}\rangle$ terms for $j=0,1,\ldots N-2, l > j$. 
    (c) The measurement circuit for estimating $\langle\hat{X}_{j}\hat{Y}_l\rangle$ for all qubit indices $j, l$ with different parity. The circuit depicted here is the special case when $N$ is even number. If $N$ is odd number, the sequence ends with the Hadamard gate on the last qubit.
    (d) Measurement gates used to rotate qubits into the common eigenbasis prior to measurement.}
    \label{fig:meas_circ}
\end{figure*}

\noindent where $\hat{X}$, $\hat{Y}$, $\hat{Z}$ are Pauli sigma matrices $\sigma_x$, $\sigma_y$, $\sigma_z$ respectively and $\hat{I}$ denotes the identity operation. For a detailed derivation of the qubit Hamiltonian, the reader is referred to the Supplementary Note 1. The diagonalisation of this Hamiltonian yields the electronic band structure, that is, the set of functions $E_{n}(\boldsymbol{k})$, where $n = 0, 1, 2, \ldots, N-1$ denotes the band index. The qubit operators $\hat{Z}_{j}$, $\hat{X}_{j}\hat{X}_{l}$, $\hat{Y}_{j}\hat{Y}_{l}$, $\hat{X}_{j}\hat{Y}_{l}$, and $\hat{Y}_{j}\hat{X}_{l}$ are Pauli $N$-qubit operators where the indices $j$ and $l$ specify the qubits on which the operations act.

\section{\textcolor{black}{Constant} Measurement Protocol}

As the Hamiltonian is represented as a sum of Pauli operators, the corresponding cost function is evaluated as the sum of their expectation values. To reduce measurement overhead in \textcolor{black}{VQE}, it is useful to group the Pauli operators appearing in Eq.~\eqref{eq.ham_qubit} into qubit-wise commuting (QWC) groups. For each QWC group, one can rotate all the qubits into the common eigenbasis and measure all operators within a QWC group simultaneously in a single round of measurement. As was done in the original work on tight-binding Hamiltonians by Sherbert \textit{et al.}~\cite{Sherbert1, Sherbert2}, and later in Ref.~\cite{Duriska}, one can group the Pauli terms into QWC groups in the following way. The first QWC group contains all the $\hat{Z}_{j}$ terms, which are measured directly in the computational basis. The second QWC group contains all the $\hat{X}_{j}\hat{X}_{l}$ terms, applying the Hadamard gate on all qubits to rotate them into the common eigenbasis. The third QWC group contains all $\hat{Y}_{j}\hat{Y}_{l}$ terms, applying the $\hat{S}^{\dagger}\hat{H}$ gates before the measurement on the computational basis. \textcolor{black}{Lastly, for a fixed $j$ and $l>j$, the terms $\hat{X}_{j}\hat{Y}_{l}$ and $\hat{Y}_{j}\hat{X}_{l}$ belong to two separate, independent QWC groups of size $N-1$}. Therefore, in general, the qubit Hamiltonian, see Eq.~\eqref{eq.ham_qubit}, contains $3 + 2(N-1) = 2N + 1$ QWC groups resulting in an asymptotic scaling of $\mathcal{O}(N)$. Although the $\mathcal{O}(N)$ scaling is efficient, we show here that the measurement overhead can be further reduced to exactly three rounds of measurements, independent of the system size.

In Supplementary Note 4, we have shown that the cost function of the qubit Hamiltonian, Eq.~\eqref{eq.ham_qubit}, can be written as

\begin{align}
    \label{eq.cost_fun}
    &E(\boldsymbol{k, \theta}) = \sum_{j = 0}^{N-1}\varepsilon_{j}\abs{a_{j}(\boldsymbol{\theta})}^2 + \sum_{j=0}^{N-2}\sum_{l>j}^{N-1}\text{Re}\{C_{jl}(\boldsymbol{\theta}) \mathcal{H}_{jl}(\boldsymbol{k})\},
\end{align}

\noindent where $\abs{a_{j}}^2$ are the probabilities of measuring the $j$-th Hamming weight 1 state (see Supplementary Note 3) and $C_{jl}$ are Pauli correlators defined as 

\begin{equation}
    \label{eq.pauli_corr}
    C_{jl}  = \langle\hat{X}_{j}\hat{X}_{l}\rangle+i\langle\hat{X}_{j}\hat{Y}_{l}\rangle = 2\abs{a_j}\abs{a_l}\mathrm{e}^{i\varphi_{jl}}.
\end{equation}

\noindent First, the probabilities $\abs{a_{j}}^2$ and hence also the absolute values of the amplitudes $\abs{a_{j}}$ can be obtained with a single round of measurement of the trial state $\ket{\psi}$ in the computational basis, see circuit $\mathcal{M}_{Z}$ in Fig.~\ref{fig:Mz}. The second step is to evaluate all correlators $C_{jl}$, for $ j = 0, 1, \ldots, N-2$, and $l > j$. Assuming all amplitudes $a_{j}\neq0$, we apply additional two measurement circuits $\mathcal{M}_{XX}$ and $\mathcal{M}_{XY}$, illustrated in Figs.~\ref{fig:Mxx},~\ref{fig:Mxy}. First, the terms $\hat{X}_{j}\hat{X}_{l}$ form one QWC group, and therefore can be estimated with one round of measurement by applying the measurement circuit $\mathcal{M}_{XX}$. The final measurement setting $\mathcal{M}_{XY}$, is implemented using a pattern of alternating XY rotations: Hadamard gates are applied to qubits with even indices (starting from 0), and the $\hat{S}^\dagger$ followed by the Hadamard gate is applied to qubits with odd indices. This configuration enables the measurement of all Pauli terms $\hat{X}_{j}\hat{Y}_{l}$ for $j$ even and $l$ such that indices $j,l$ have different parity $\mathcal{P}_{\text{diff}}$ and $\hat{Y}_{j}\hat{X}_{l}$ for $j$ odd and $l$ such that indices $j,l \in \mathcal{P}_{\text{diff}}$ because they form one QWC group. 

Note that the $\hat{Y}_{j}\hat{X}_{l}$ terms are not needed in evaluating the Pauli correlator, see Eq.~\eqref{eq.pauli_corr}. However, the terms $\hat{X}_{j}\hat{Y}_{l}$ for $j$ odd $l$ such that $j,l \in \mathcal{P}_{\text{diff}}$ are required. As shown in the Supplementary Note 4, within the single-electron approximation, the expectation values satisfy the relation $\langle \hat{X}_{j}\hat{Y}_{l} \rangle = -\langle \hat{Y}_{j}\hat{X}_{l} \rangle$. This means that the necessary terms $\hat{X}_{j}\hat{Y}_{l}$ for $j$ odd and $j,l \in \mathcal{P}_{\text{diff}}$ can still be extracted from the same measurement setting by simply flipping the sign, even though they are not part of the original QWC group. Therefore, the measurement circuit $\mathcal{M}_{XY}$ enables us to measure all Pauli correlators $C_{jl}$ for qubit indices $j,l$ with different parity.

There are still many Pauli correlators left in Eq.~\eqref{eq.cost_fun}, namely $C_{jl}$, for qubit pairs $j, l$ with the same parity $\mathcal{P}_{\text{same}}$, that are not directly obtained from the three measurement settings described above. However, no additional measurement is needed because they can be reconstructed from the measurement statistics obtained from the previous measurements using the product rule formula

\begin{equation}
    \label{eq.prod_rule}
    C_{jl} = \dfrac{C_{jk}C_{kl}}{2\abs{a_{k}}^2},\quad \text{for $j,l \in \mathcal{P}_{\text{same}}$}
\end{equation}

\noindent where the index $k$ is chosen such that $j,k \in \mathcal{P}_{\text{diff}}$ and $k,l \in \mathcal{P}_{\text{diff}}$. Notably, the estimation of the entire cost function can be achieved with a constant three rounds of measurement, regardless of the size of the system.

\begin{figure*}[t!]  
    \centering
    \begin{subfigure}{0.49\textwidth} 
        \centering
        \includegraphics[width=\linewidth]{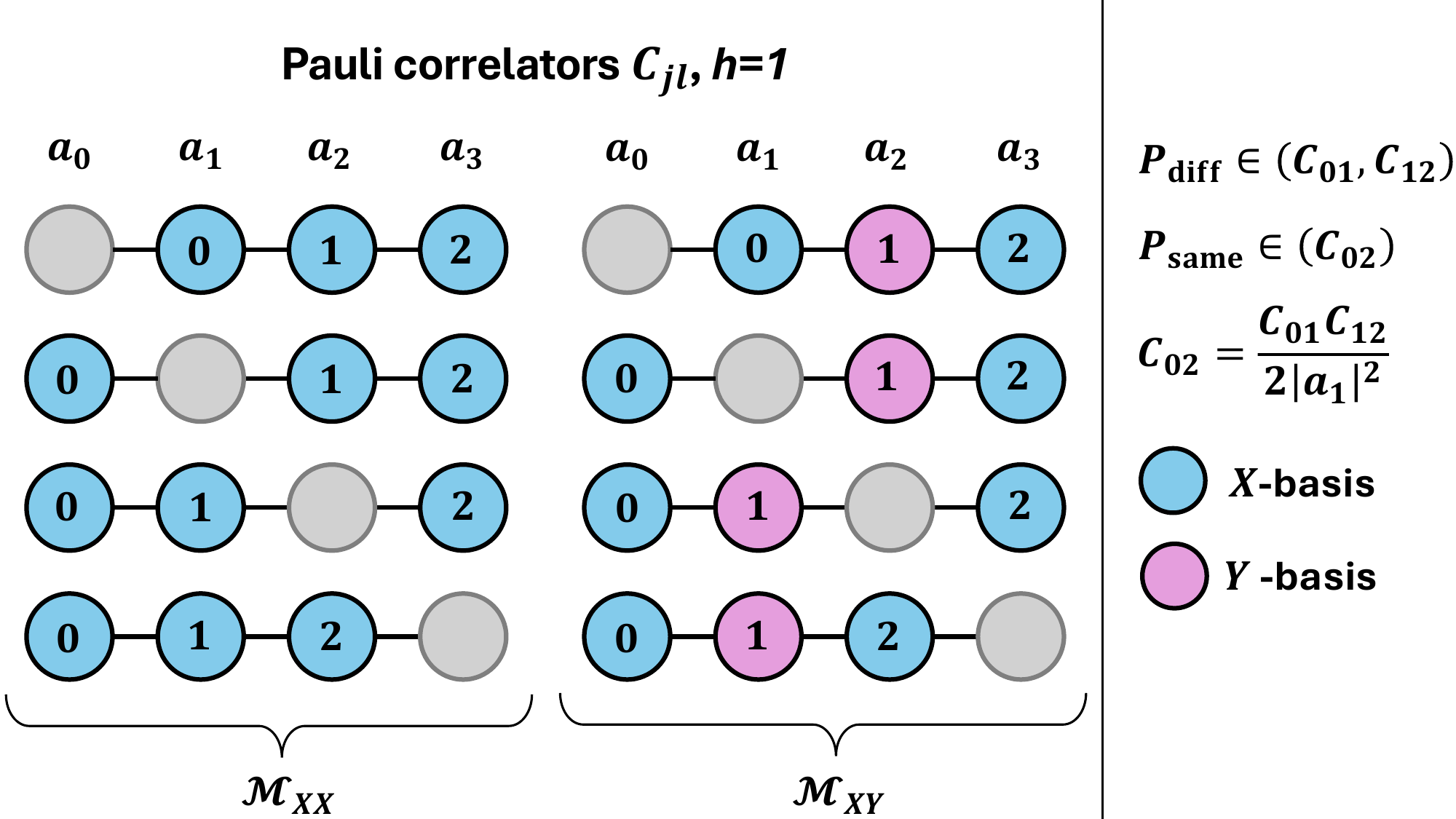} 
        \caption{}
        \label{fig.zero_amps}
    \end{subfigure}
    \hfill
    \begin{subfigure}{0.49\textwidth} 
        \centering
        \includegraphics[width=\linewidth]{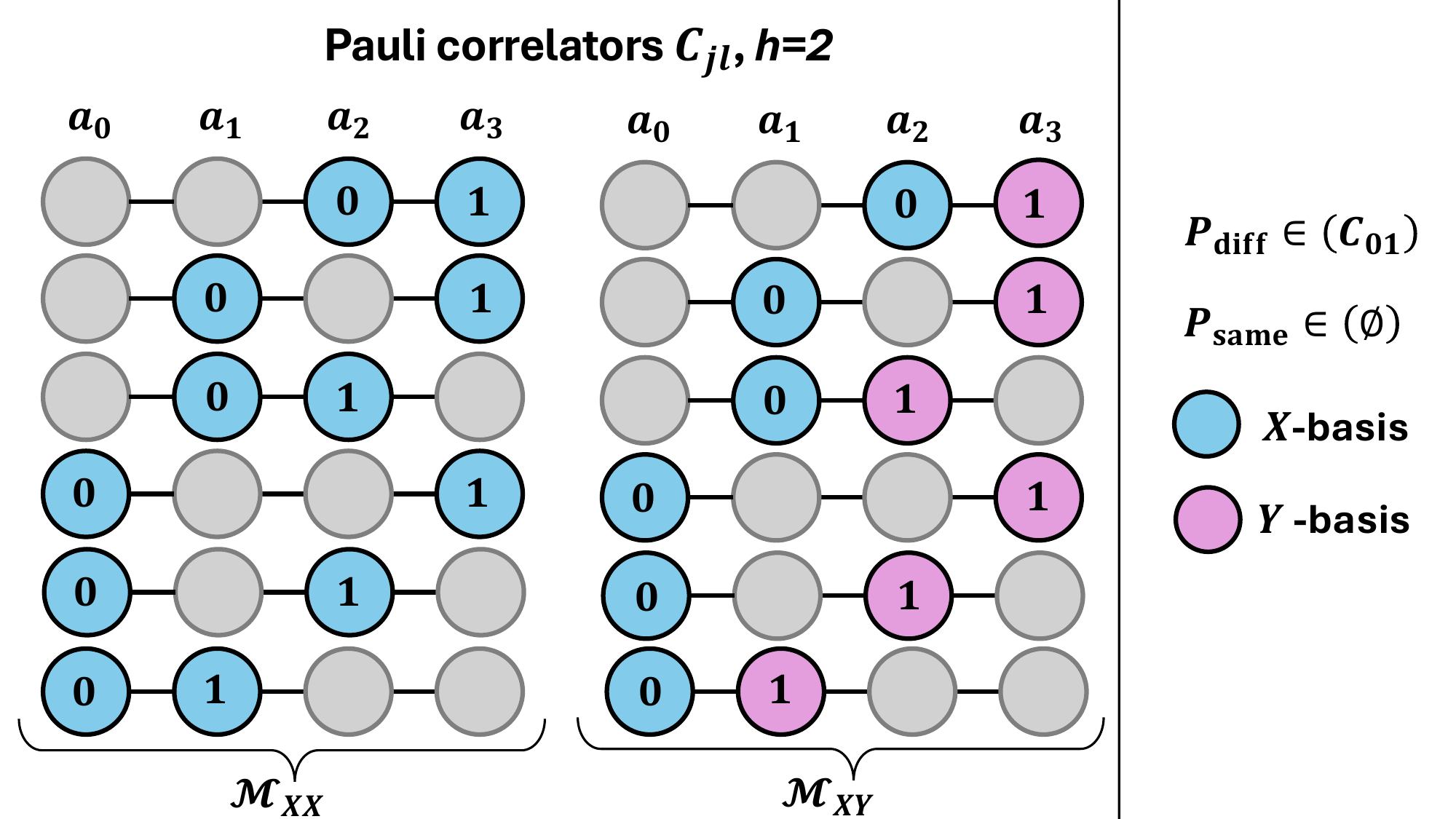}
        \caption{}
        \label{fig.zero_amps1}
    \end{subfigure}
    \caption{Example of the measurement strategy for the 4-qubit model. Grey indicates zero amplitudes, while blue and purple denote the application of measurement gates that rotate the qubits into the common eigenbasis: $X$ and $Y$, respectively. (a) All possible situations for zero amplitude (grey color) events for $h = 1$. The corresponding Pauli correlators that contribute to the cost function, see Eq.~\eqref{eq.cost_fun}, are obtained directly from the measurements ($\mathcal{P}_{\text{diff}}$) and indirectly using the product rule formula~\eqref{eq.prod_rule} ($\mathcal{P}_{\text{same}}$). (b) All possible situations for zero amplitude (grey color) events for $h = 2$. The corresponding Pauli correlators that contribute to the cost function, see Eq.~\eqref{eq.cost_fun}, are obtained directly from the measurements ($\mathcal{P}_{\text{diff}}$). There are no indirect Pauli correlators needed. }
    \label{fig.zero_amplitudes}
\end{figure*}

The complication can occur when the assumption $a_{j} \neq 0 $ fails. During the optimisation, this can occur either when the optimiser proposes a set of angles \(\boldsymbol{\theta}\) such that some amplitudes $a_{j}$ are exactly zero, or for basis states that have small but nonzero amplitudes which are not observed due to finite shot count; the latter are referred to as apparent zeros. In the cases where $a_{j} = 0 $ for some indices $j$, we can still use the three rounds of the measurement protocol as described above simply by ignoring the qubits with zero amplitudes. First, we identify any amplitudes $a_{j}$ that vanish. Since the cost function, Eq.~\eqref{eq.cost_fun} includes terms of the form $\abs{a_j}\abs{a_l}$, if either $a_j$ or $a_l$ is zero, the contribution of that term vanishes and may be ignored. Let $h$ denote the number of zero amplitudes. We then define the compressed index set $S = \{s_0 < s_1 < \cdots < s_{m-1}\}$, corresponding to the qubits with nonzero amplitudes $a_{s_k} \neq 0$, where $m = N - h$ and $N$ is the total number of qubits. This list is obtained by discarding the zero-amplitude qubits and reindexing the rest starting from $0$. If $h=0$, the compressed set $S$ coincides with the original index set. After this step, we proceed as described above with the measurement circuits $\mathcal{M}_{XX}$ and $\mathcal{M}_{XY}$ only now applied to the reindexed qubits, see Fig.~\ref{fig.zero_amplitudes} that illustrate the strategy of applying the measurement circuits $\mathcal{M}_{XX}$, $\mathcal{M}_{XY}$ for the case of 4-qubit model with $h = 1, 2$. Figure~\ref{fig.zero_amps} shows all possible situations where one of the amplitudes $\abs{a_{j}}$ for $j = 0, 1, 2, 3$ is zero. On the other hand, Fig.~\ref{fig.zero_amps1} shows six possible cases where two of the amplitudes are zero. The case for $h=3, 4$ is trivial because in that case the second term in Eq.~\eqref{eq.cost_fun} does not contribute to the cost function.

\begin{figure*}[htpb]
    \centering
    \includegraphics[width=0.99\textwidth]{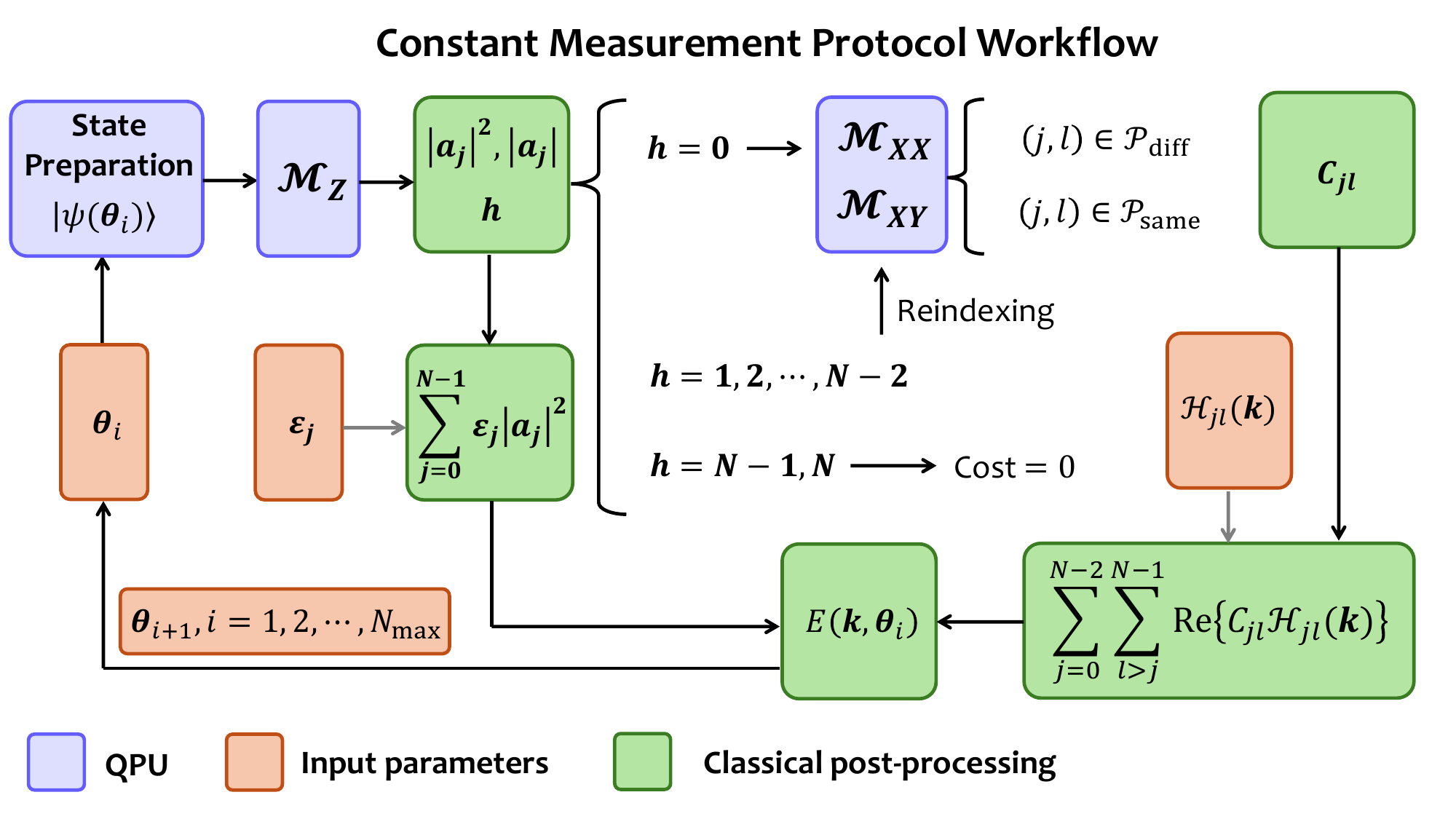}  
    \caption{Workflow of the constant measurement protocol. The QPU denotes the parts of the algorithm that are carried out by a quantum computer or simulator. The input parameters are the variational angles $\boldsymbol{\theta}_{i}$ and the model parameters  $\varepsilon_{j}$, $\mathcal{H}(\boldsymbol{k})_{jl}$. $N_{\text{max}}$ denotes the maximum number of iterations. The symbol $h$ represents the number of zero amplitudes. The green shaded parts depict the classical post-processing.}
    \label{fig:vqd_workflow}
\end{figure*}

In summary, the workflow of the constant measurement protocol is illustrated in Fig.~\ref{fig:vqd_workflow} and can be summarized as follows:

\begin{itemize}
  \item \(\boxed{\mathcal{M}_{Z}}\)  
  A single round of measurements in the computational basis yields $\abs{a_{j}}^2$, and hence $\abs{a_{j}}$, for $j = 0, 1, \ldots, N-1$. In this step, we identify any amplitudes $\abs{a_{j}}$ that vanish. 

  \item \(\boxed{\mathcal{M}_{XX}}\)  
  Based on the number of zero amplitudes, we apply the measurement circuit $\mathcal{M}_{XX}$, yielding the expectation values $\langle \hat{X}_{j} \hat{X}_{l} \rangle$ for all pairs with $j = 0, 1, \ldots, m-2$ and $l > j$. 

  \item \(\boxed{\mathcal{M}_{XY}}\)  
  Similarly, the application of the measurement circuit $\mathcal{M}_{XY}$ yields the expectation values $\langle \hat{X}_{j} \hat{Y}_{l} \rangle$ for pairs of qubits $j = 0, 1, \ldots, m-2$ and $l > j$ with different parity $\mathcal{P}_{\text{diff}}$. Combining these results with those obtained from $\mathcal{M}_{XX}$ allows us to determine the corresponding Pauli correlators $C_{jl}$ using formula~\eqref{eq.pauli_corr}. For pairs of qubits $j, l \in \mathcal{P}_{\text{same}}$, the Pauli correlators can be reconstructed using the product rule formula~\eqref{eq.prod_rule}.
\end{itemize}

\begin{figure}[t!]
    \centering
    \includegraphics[width=\columnwidth]{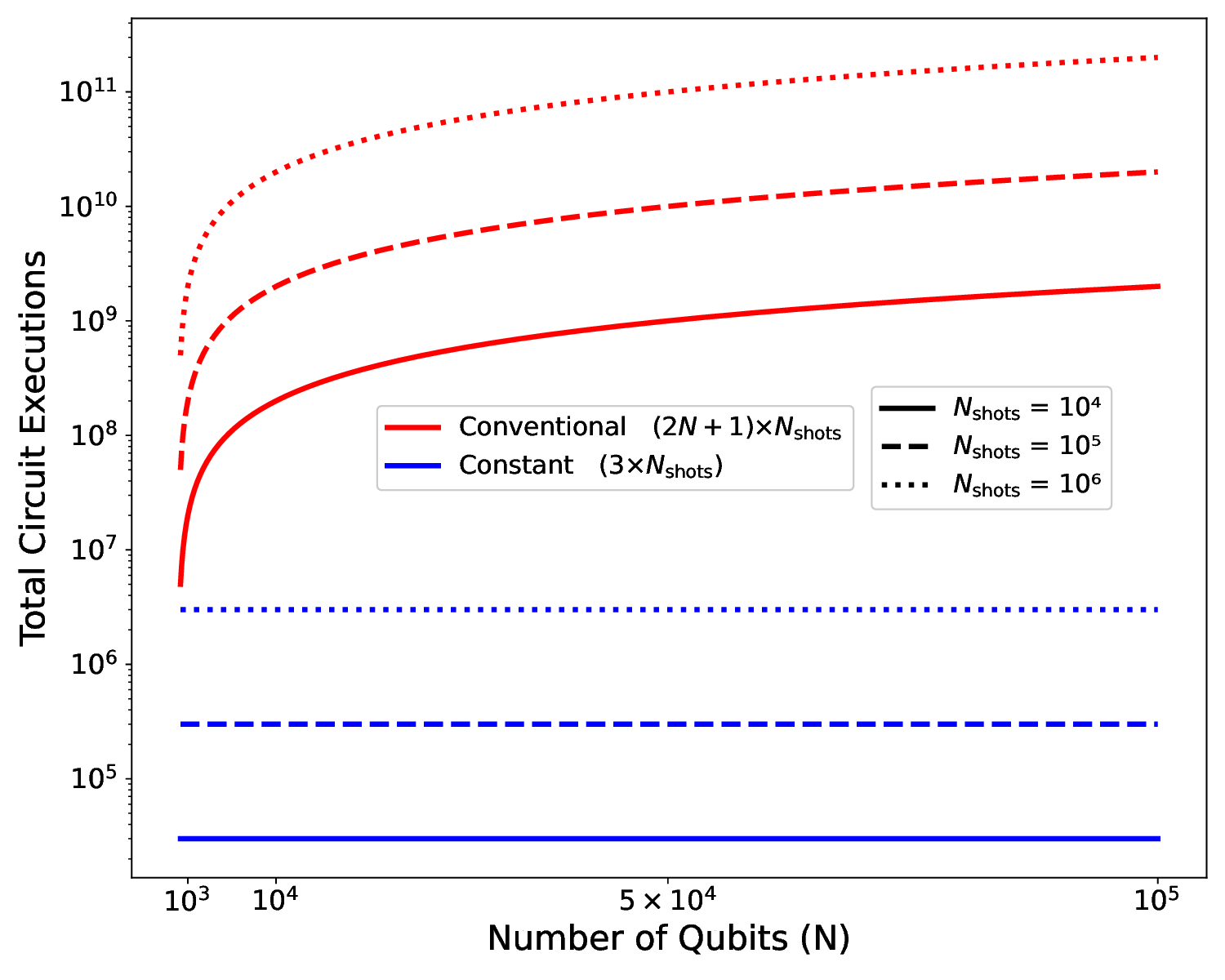} 
    \caption{Total number of circuit executions (log scale) for the cost function estimation, comparing conventional and constant measurement protocols with $N_{\text{shots}} = 10^4, 10^5, 10^6$.}
    \label{fig:circuit_executions}
\end{figure}

\noindent Each measurement circuits $\mathcal{M}_{Z}$, $\mathcal{M}_{XX}$, and $\mathcal{M}_{XY}$ is executed $N_{\text{shots}}$ times. The Figure~\ref{fig:circuit_executions} compares the conventional $\mathcal{O}(N)$ measurement protocol with our constant-depth approach. As this figure shows, the advantage of the constant protocol becomes increasingly pronounced for larger systems. While the number of measurement settings in the conventional method scales linearly with system size, this still leads to a significant increase in total quantum overhead — especially when high shot counts are required for statistical accuracy. For large models, the constant protocol becomes essential for maintaining scalability on near-term quantum hardware.

\section{Results and Discussion}\label{sec4}

In this section, we present several benchmark calculations using the \textcolor{black}{OA-VQE} algorithm with a constant $\mathcal{O}(1)$ measurement protocol and compare them against results obtained via exact diagonalisation. We report results for two tight-binding models: a three-qubit model of a two-dimensional CuO$_2$ square lattice with a three-atom basis~\cite{Fulde1995}, and a four-qubit model of a two-dimensional bilayer graphene system~\cite{McCann, Kuzmenko}. \textcolor{black}{These models were evaluated using a shot-based simulator implemented within the open-source quantum computing framework Qiskit~\cite{Qiskit2024}, as well as on \textcolor{black}{IBM} physical quantum hardware. For both the simulator and hardware runs, classical parameter optimisation was performed using the COBYQA (Constrained Optimisation BY Quadratic Approximation) algorithm~\cite{Ragonneau} as implemented in SciPy~\cite{SciPy}.}

\begin{figure*}[t]
    \centering
    \begin{subfigure}[t]{0.49\textwidth}
        \centering
    \includegraphics[width=\linewidth]{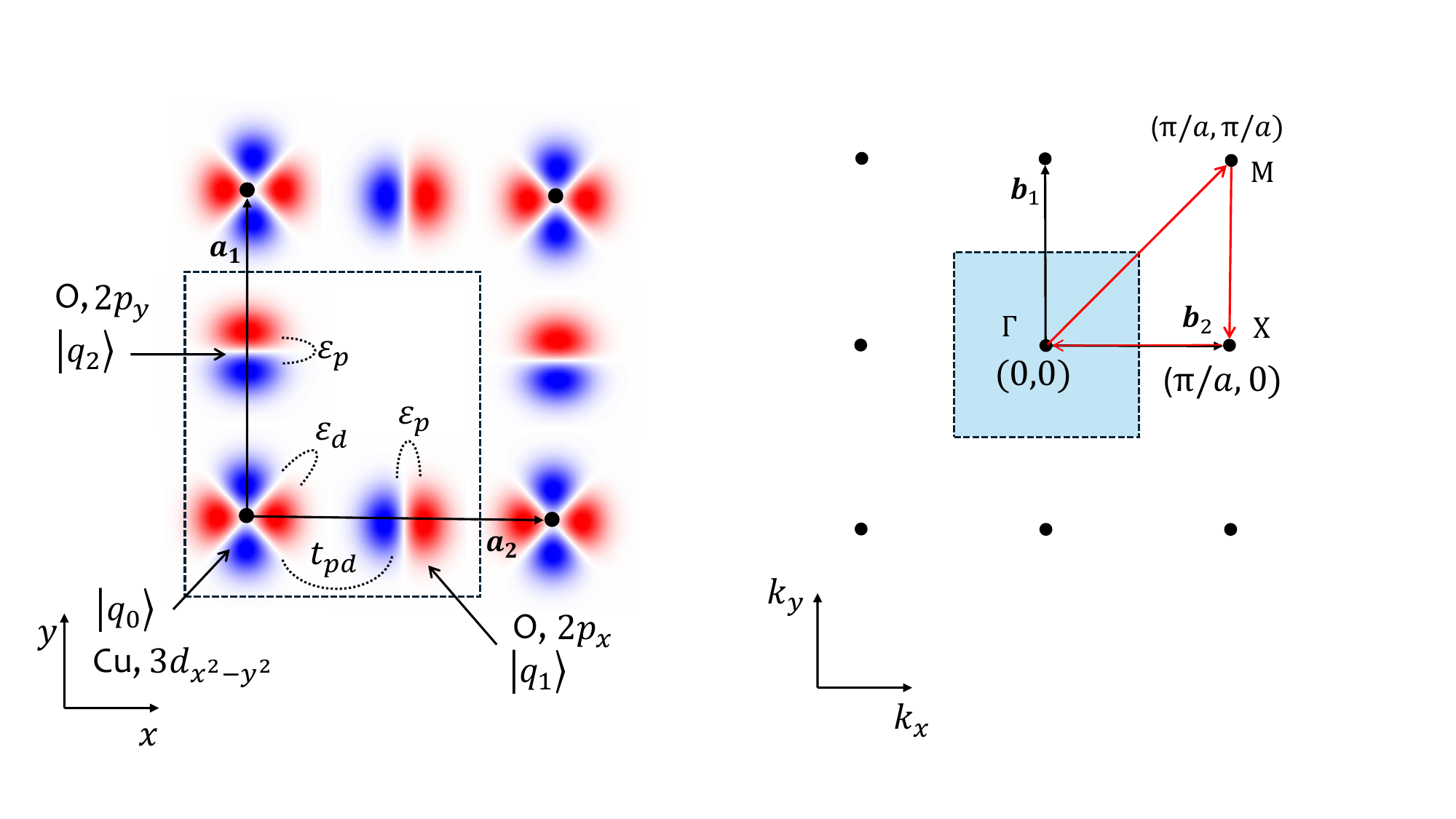}
        \caption{\quad \quad \quad \quad \quad \quad \quad \quad \quad \quad \quad \quad (b)}
    \end{subfigure}
    \hfill
    \begin{subfigure}[t]{0.49\textwidth}
        \centering
        \includegraphics[width=\linewidth]{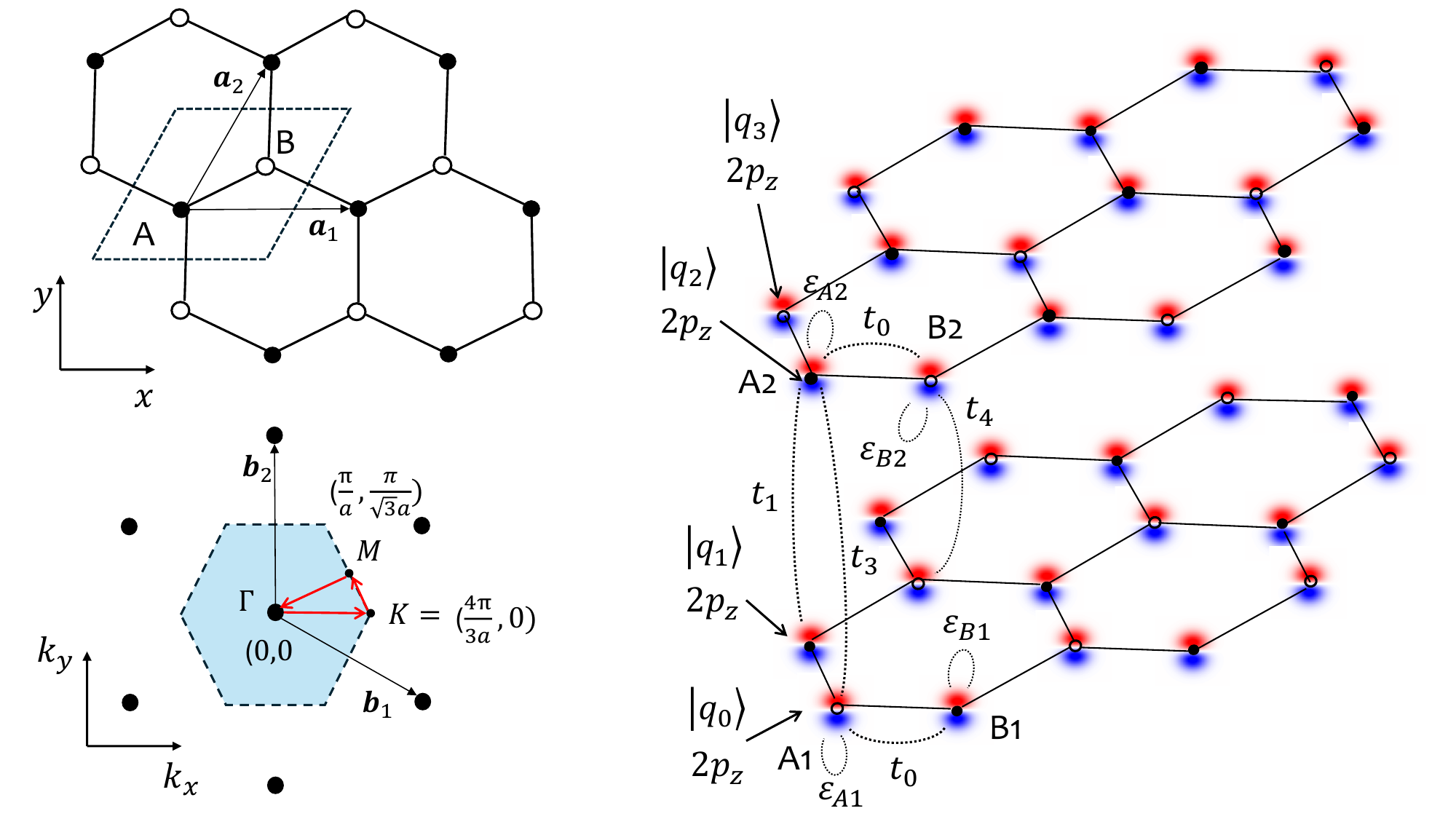}
        (d)\quad \quad \quad \quad \quad \quad \quad \quad \quad \quad \quad \quad (e)
    \end{subfigure}

    
    \begin{subfigure}[t]{0.49\textwidth}
        \centering
        \includegraphics[width=\linewidth]{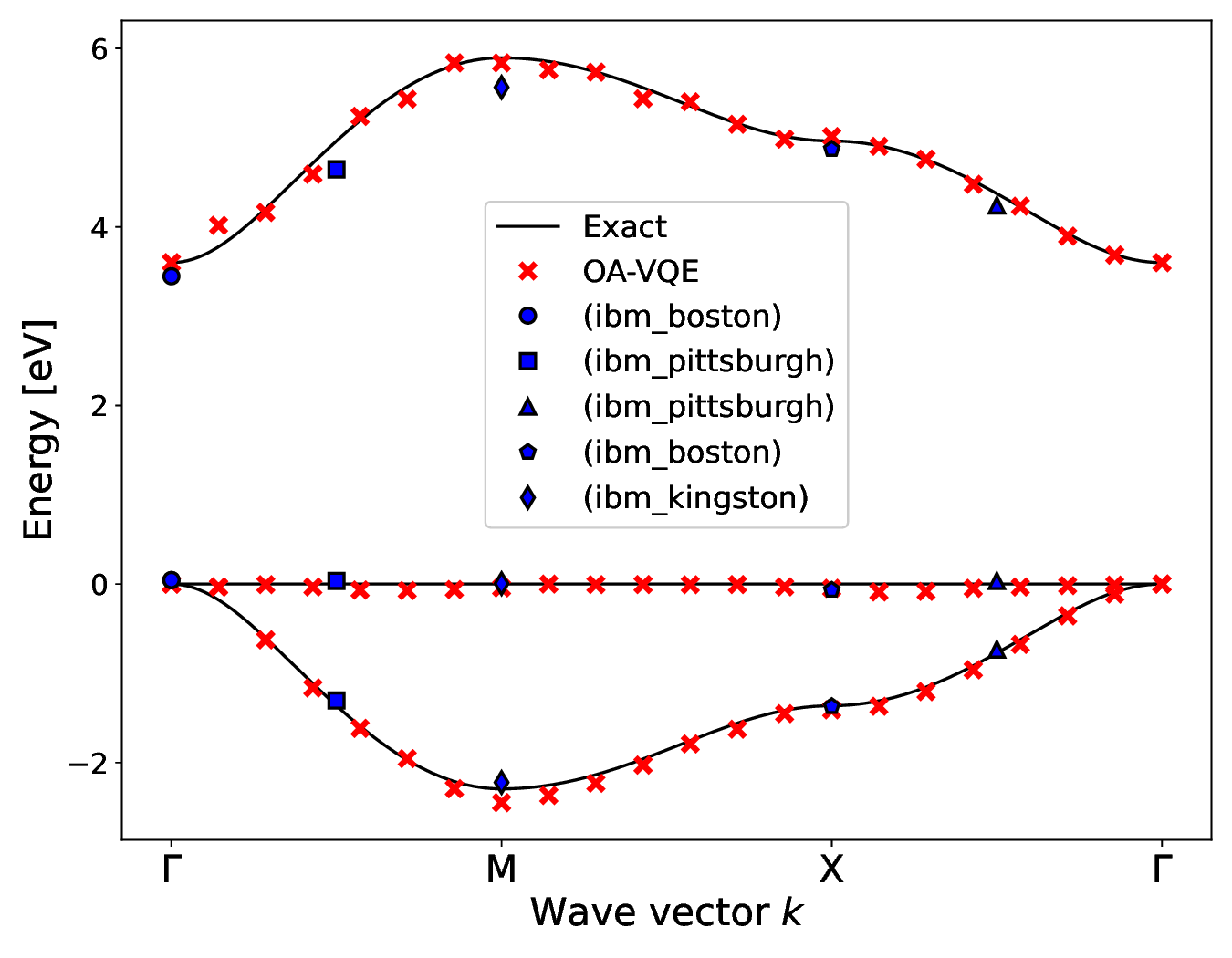}
        (c)
    \end{subfigure}
    \hfill
    \begin{subfigure}[t]{0.49\textwidth}
        \centering
        \includegraphics[width=\linewidth]{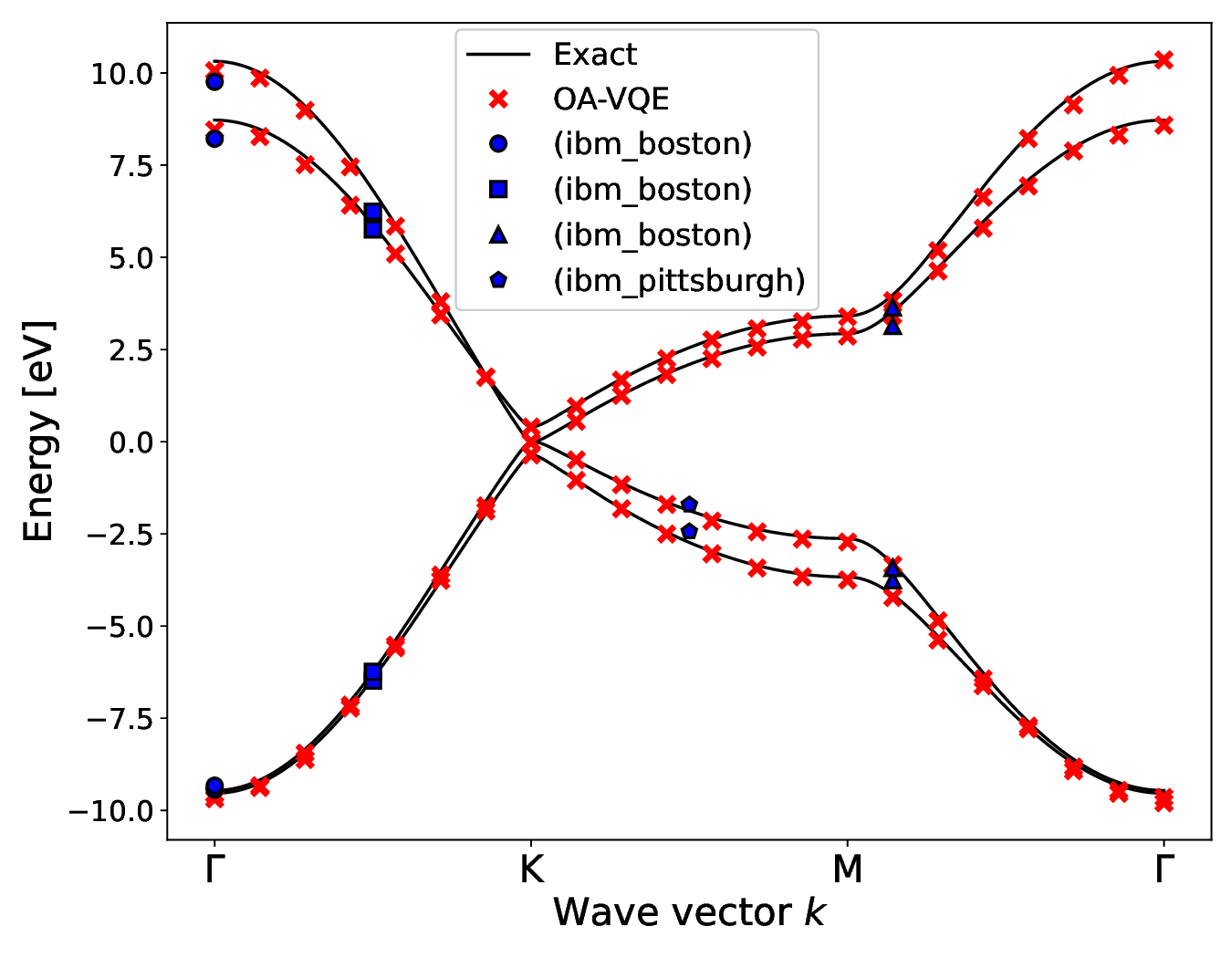}
        (f)
    \end{subfigure}

    \caption{Comparison of two models. (a) CuO$_2$ structure - unit cell with three atomic orbitals and the relevant hopping amplitudes. (b) The first Brillouin zone of CuO$_2$ and the high-symmetry path. (c) \textcolor{black}{Three-qubit model CuO$_2$ band structure} (\textcolor{black}{black} lines: exact diagonalisation, red crosses: \textcolor{black}{OA-VQE simulator} results; \textcolor{black}{blue symbols: real hardware results}). (d) Bilayer graphene structure - Monolayer top view and the first Brillouin zone with the high-symmetry path. (e) Bilayer graphene side view with relevant hopping amplitudes. (f) \textcolor{black}{Four-quibit model bilayer graphene band structure (black lines: exact diagonalisation, red crosses: OA-VQE results, blue symbols: real hardware results ).}}
    \label{fig:combined_2x2}
\end{figure*}

In addition, to further validate the scalability of our measurement protocol, we benchmarked the \textcolor{black}{cost function~(\ref{eq.cost_fun})} estimation \textcolor{black}{for three, four, and ten qubit models.} Importantly, this benchmarking is performed \emph{independently of the variational optimisation}, which becomes increasingly difficult and unreliable for larger systems due to the growing complexity of the cost landscape and \textcolor{black}{the compounding statistical error arising from the increased number of measurable terms in the Hamiltonian}. By decoupling the protocol from classical optimisation and focusing purely on the quantum estimation of \textcolor{black}{the cost function}, we demonstrate that our method remains effective and scalable even when applied to system sizes at or near the practical limits of variational quantum algorithms under realistic noise conditions. We find that the \textcolor{black}{estimated} values of \textcolor{black}{the cost function} remain accurate and exhibit stable variance across system sizes, confirming that the measurement cost and precision do not deteriorate with qubit count.

\subsection{Benchmark calculations of the constant  measurement protocol}
Figure~\ref{fig:combined_2x2} summarises benchmark calculations for representative tight-binding models. The three- and four-qubit models represent the CuO$_2$ plane and bilayer graphene, respectively. The numerical values for the on-site energies and hopping amplitudes in the three- and four-qubit models were taken from Refs.~\cite{Fulde1995, McCann, Kuzmenko}. Solid \textcolor{black}{black} curves \textcolor{black}{depict} the results obtained via exact diagonalisation, while red crosses indicate the energies computed using our constant measurement protocol with the warm-start optimisation strategy \textcolor{black}{implemented on the shot-based simulator}. \textcolor{black}{The blue symbols represent the experimental energies evaluated on physical superconducting quantum hardware. For both the simulated and real hardware environments, a fixed measurement budget of $N_{\text{shots}} = 8192$ shots per setting was used}. \textcolor{black}{To preserve quantum state coherence and protect the circuits against environmental noise during physical execution, active error suppression and mitigation strategies were deployed via the Qiskit \emph{Estimator} primitive. Specifically, Twirled Readout Error Extrapolation (TREX) was utilised for measurement mitigation (\emph{Resilience Level 1}) alongside an XY4 Dynamical Decoupling (DD) sequence with gate twirling enabled to counteract cross-talk and idle qubit decoherence. Across all configurations, the simulated and hardware-enacted results exhibit excellent agreement with the analytical curves, demonstrating that the proposed protocol accurately reproduces the expected spectra while remaining highly robust against both finite shot noise and hardware systematic errors.}

\subsection{\textcolor{black}{Statistical Analysis of Cost Function Estimation}}

\textcolor{black}{Cost function estimation~(\ref{eq.cost_fun}) with just three measurement bases relies primarily on the product rule, Eq.~(\ref{eq.prod_rule}) to extract the Pauli correlators $C_{jl}$ with the same parity indices $j,l \in \mathcal{P}_{\text{same}}$. Since the correlators $C_{jl}$ for $j,l \in \mathcal{P}_{\text{diff}}$ are extracted from the same measurement data and the correlators $C_{jl}$ for $j,l \in P_{\text{same}}$ are further constructed with the product rule, which is a non-linear function of other estimated quantities, the results are statistically correlated. To estimate the variance and statistical stability of the constant measurement protocol, we performed 50 independent trials of cost function estimation at the exact energy eigenstates points across the entire band structure and computed the sample mean $\bar{E}_{n}(\boldsymbol{k})$ and standard deviation $\sigma_{n}(\boldsymbol{k})$ as }

\begin{equation}
    \label{eq.mean}
    \bar{E}_{n}(\boldsymbol{k}) = \dfrac{1}{M} \sum_{i=1}^{M} E_{n}(\boldsymbol{k})^{(i)},
\end{equation}

\begin{equation}
    \sigma_{n}(\boldsymbol{k}) = \sqrt{\frac{1}{M-1} \sum_{i=1}^{M} \left[E_{n}(\boldsymbol{k})^{(i)} - \bar{E}_{n}(\boldsymbol{k})\right]^2}.
\end{equation}

 \noindent To eliminate errors originating from the classical optimisation, we precomputed the variational angles using an exact statevector estimator, which yields the exact angles for each eigenstate. These experiments were performed for both our three- and four-qubit models. Additionally, we benchmarked the protocol in the 10-qubit regime using the three-dimensional diamond \textcolor{black}{silicon} tight-binding Hamiltonian investigated in our previous work \cite{Duriska}. We executed these experiments for both the $\mathcal{O}(1)$ constant measurement protocol and the standard $\mathcal{O}(N)$ protocol, calculating the variance ratio $R_{\sigma^2}$ at each $\boldsymbol{k}$-point for each state $n = 0, 1, 2, \ldots, N-1$.
 
 \begin{equation}
     \label{eq.var_ratio}
     R_{\sigma^2} = \dfrac{\sigma_{\mathcal{O}(1)}^{2}}{\sigma_{\mathcal{O}(N)}^2}.
 \end{equation}
 
\begin{figure*}[t]
    \centering
    \begin{subfigure}[t]{0.49\textwidth}
        \centering
    \includegraphics[width=\linewidth]{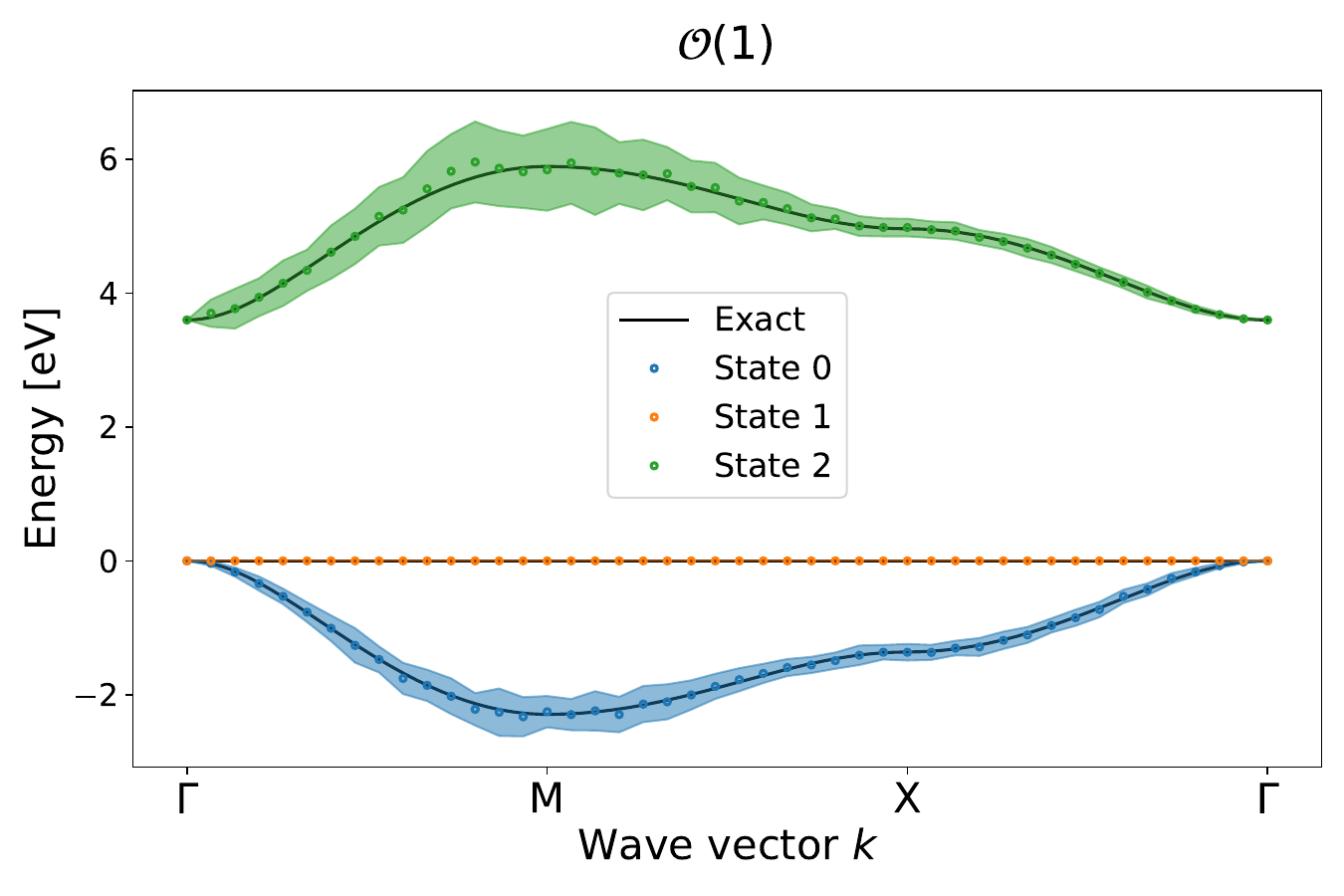}
        \caption{}
    \end{subfigure}
    \hfill
    \begin{subfigure}[t]{0.49\textwidth}
        \centering
        \includegraphics[width=\linewidth]{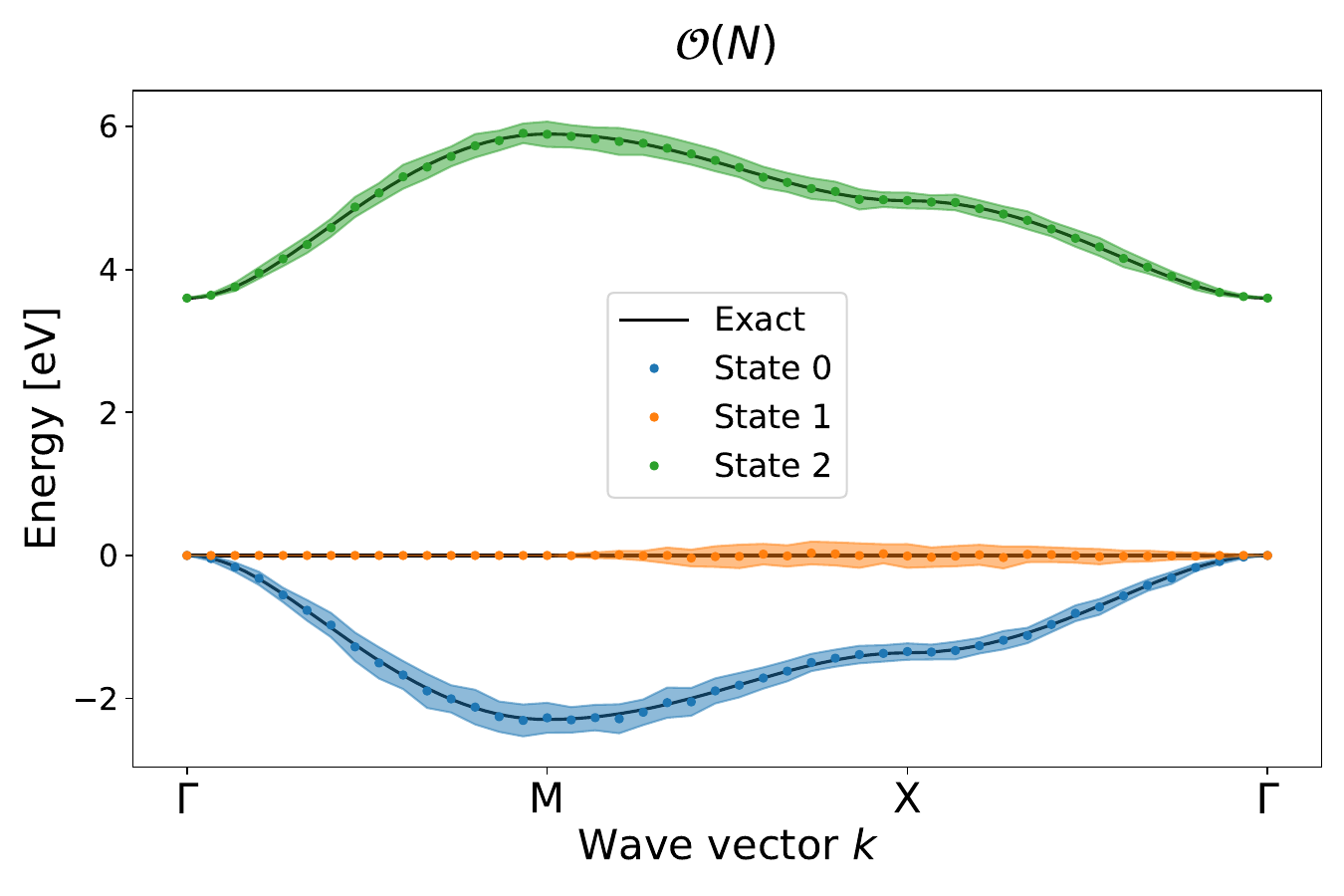}
        (b)
    \end{subfigure}

    \begin{subfigure}[t]{0.49\textwidth}
        \centering
        \includegraphics[width=\linewidth]{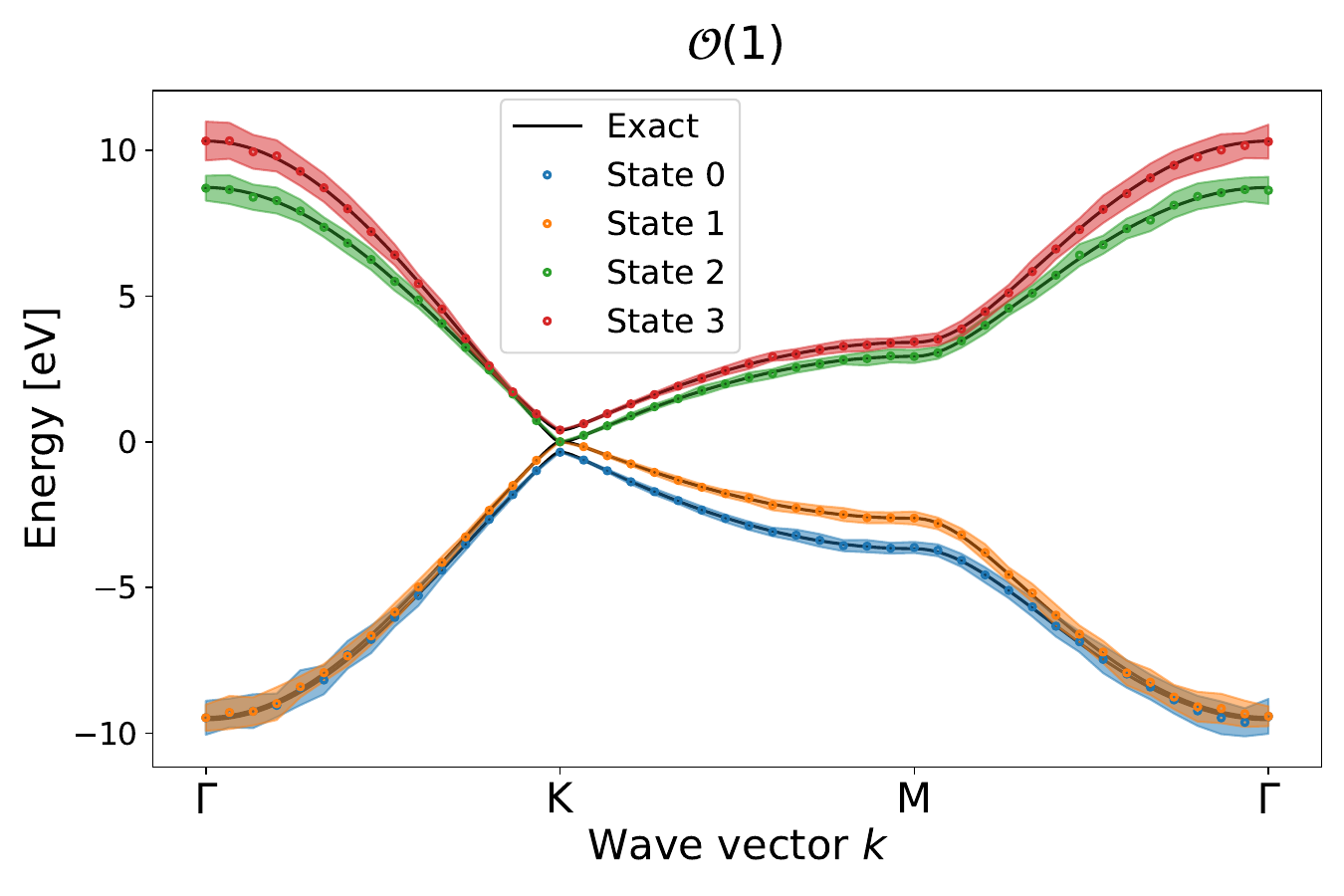}
        (c)
    \end{subfigure}
    \hfill
    \begin{subfigure}[t]{0.49\textwidth}
        \centering
        \includegraphics[width=\linewidth]{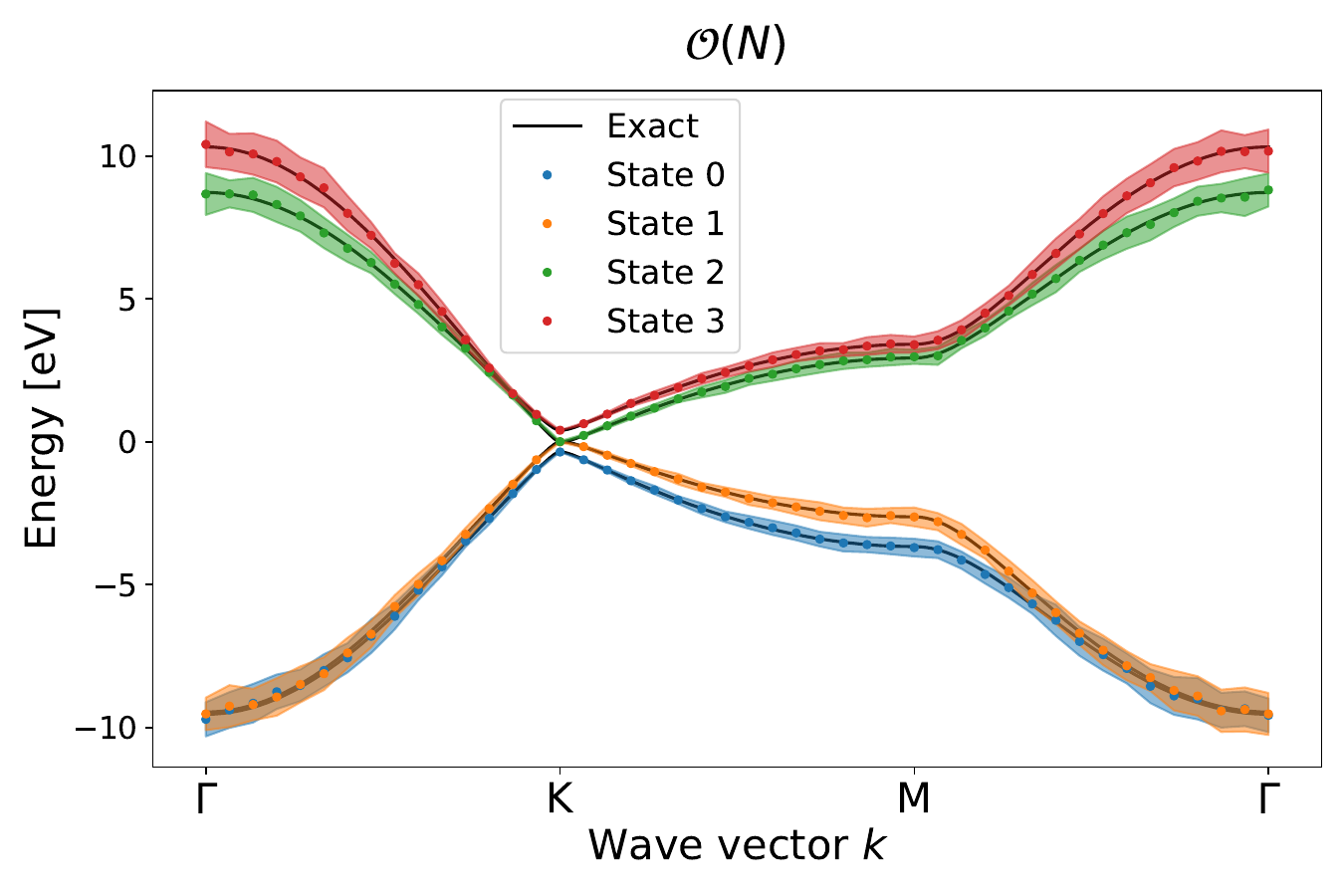}
        (d)
    \end{subfigure}

    \begin{subfigure}[t]{0.49\textwidth}
        \centering
        \includegraphics[width=\linewidth]{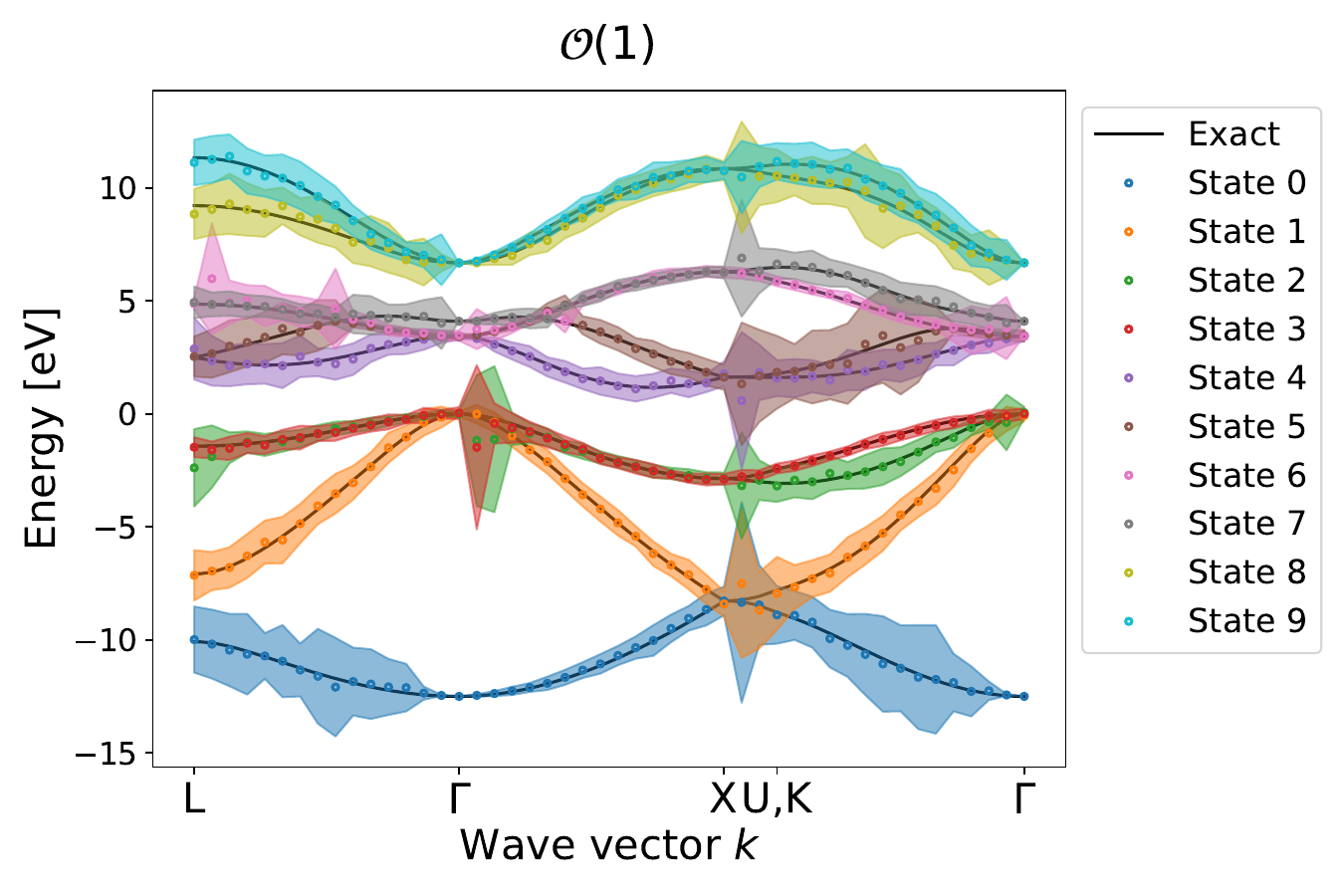}
        (e)
    \end{subfigure}
    \hfill
    \begin{subfigure}[t]{0.49\textwidth}
        \centering
        \includegraphics[width=\linewidth]{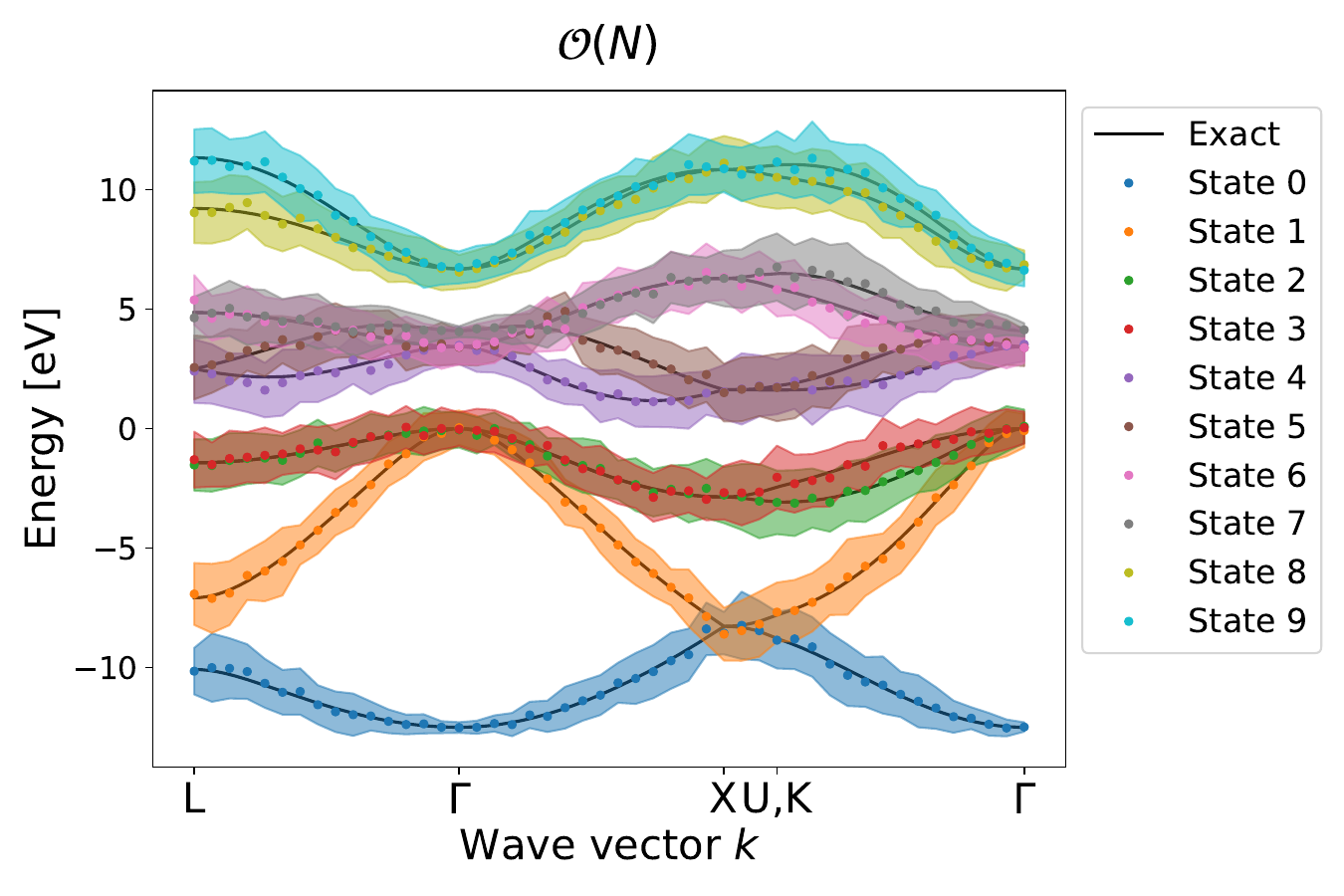}
        (f)
    \end{subfigure}

    \caption{\textcolor{black}{Standard deviations of the cost function computed with the constant $\mathcal{O}(1)$ protocol are shown in the left column, and those computed with the standard $\mathcal{O}(N)$ protocol are on the right. In all cases, the number of shots was set to $N_{\text{shots}}=10^3$}.}
    \label{fig:statistics}
\end{figure*}

 In the Figure \ref{fig:statistics} we show the band structures for our three-, four-, and ten-qubit models, along with the calculated mean energies $\bar{E}_{n}(\boldsymbol{k})$ and their standard deviations $\sigma_{n}(\boldsymbol{k})$, which are depicted as shaded regions around the means. The experiments were performed using a fixed total budget of $N_{\text{shots}} = 10^{3}$ shots. In the Supplementary Note 5, there are additional data for $N_{\text{shots}} = 10^{4}$ shots. The constant measurement protocol yields better precision in the cost function estimation for larger systems, an advantage that becomes more pronounced as the system size grows. This behaviour arises because the constant measurement protocol only requires $N_{\text{shots}}$ to be split among three measurement bases, whereas the standard protocol must distribute this budget across $\mathcal{O}(N)$ bases. As the qubit number $N$ scales, the constant measurement protocol allocates significantly more shots per measurement basis than the conventional protocol. For the three-qubit CuO$_2$ model, which is the special case because both protocols require exactly three measurement bases, the conventional protocol, $\mathcal{O}(N)$, exhibits better overall precision, except for the first excited state along the M-X-$\Gamma$ path. In the four and ten-qubit models, the overall precision appears to be better for the constant measurement protocol. However, in the case of the ten-qubit model, there remain regions in the Brillouin zone, specifically around the $\Gamma$, X, U, and K high-symmetry points, where the constant measurement protocol appears unstable and exhibits high variance. This behaviour is an artifact of the nonlinearity of the product rule combined with the larger numerical magnitudes of the off-diagonal coefficients in the Hamiltonian. In these regimes, the constant measurement protocol struggles to accurately resolve the energy due to noise amplification. In special cases where the expectation values of the Pauli correlators $C_{jl}$ are non-negligible, and the absolute values of the off-diagonal elements in the Hamiltonian $\abs{\mathcal{H}_{jl}}$ are simultaneously large, the nonlinear nature of the product rule drives significant noise amplification. 
 
\begin{figure}[htpb]
    \centering
    \includegraphics[width=0.49\textwidth]{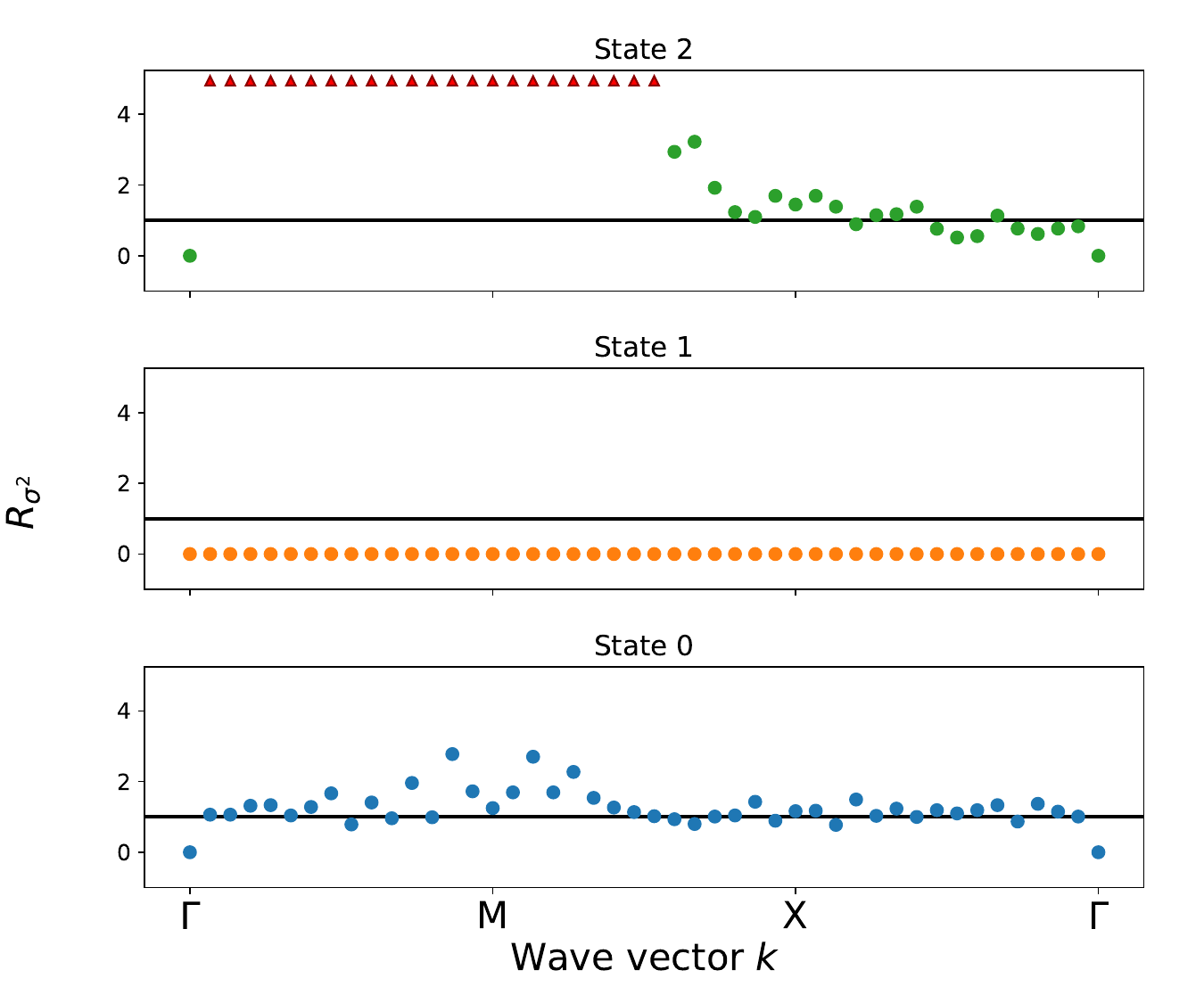} 
    \caption{\textcolor{black}{Variance ratios $R_{\sigma^2}$ for the CuO$_2$ three qubit model as a function of wave vector $\boldsymbol{k}$ for each eigenstate. The red triangles denote the outlier ratios where $R_{\sigma^2} > 5$. The horizontal solid black line corresponds to the value of 1, values under this threshold indicate a better performance by the proposed constant-measurement protocol.}}
    \label{fig.R_CuO2}
\end{figure}

\begin{figure}[htpb]
    \centering
    \includegraphics[width=0.49\textwidth]{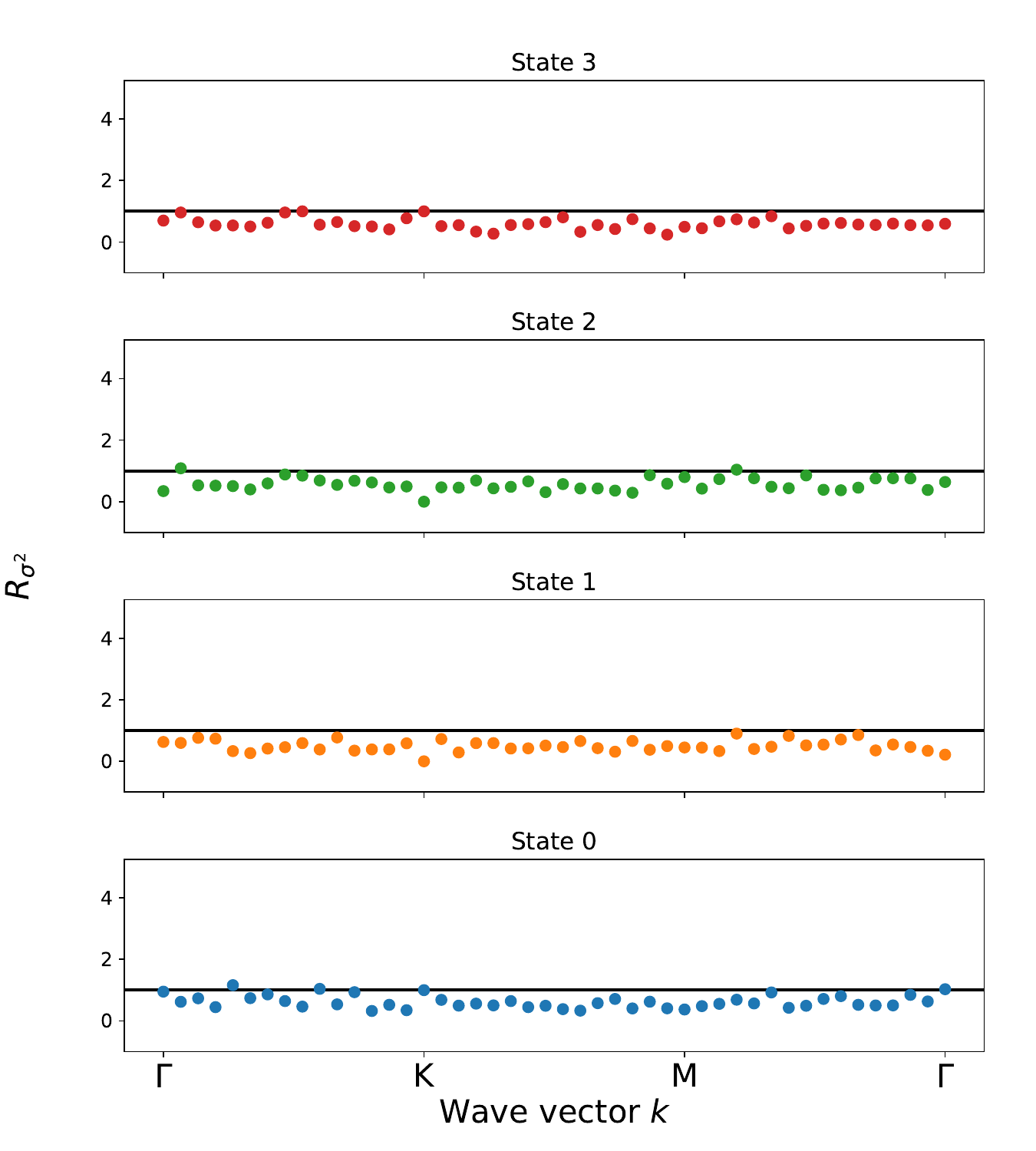} 
    \caption{\textcolor{black}{Variance ratios $R_{\sigma^2}$ for the four-qubit bilayer graphene model as a function of wave vector $\boldsymbol{k}$ for each eigenstate. The horizontal solid black line corresponds to the value of 1, values under this threshold indicate a better performance by the proposed constant-measurement protocol.}}
    \label{fig.R_bg}
\end{figure}

\begin{figure}[htpb]
    \centering
    \includegraphics[width=0.49\textwidth]{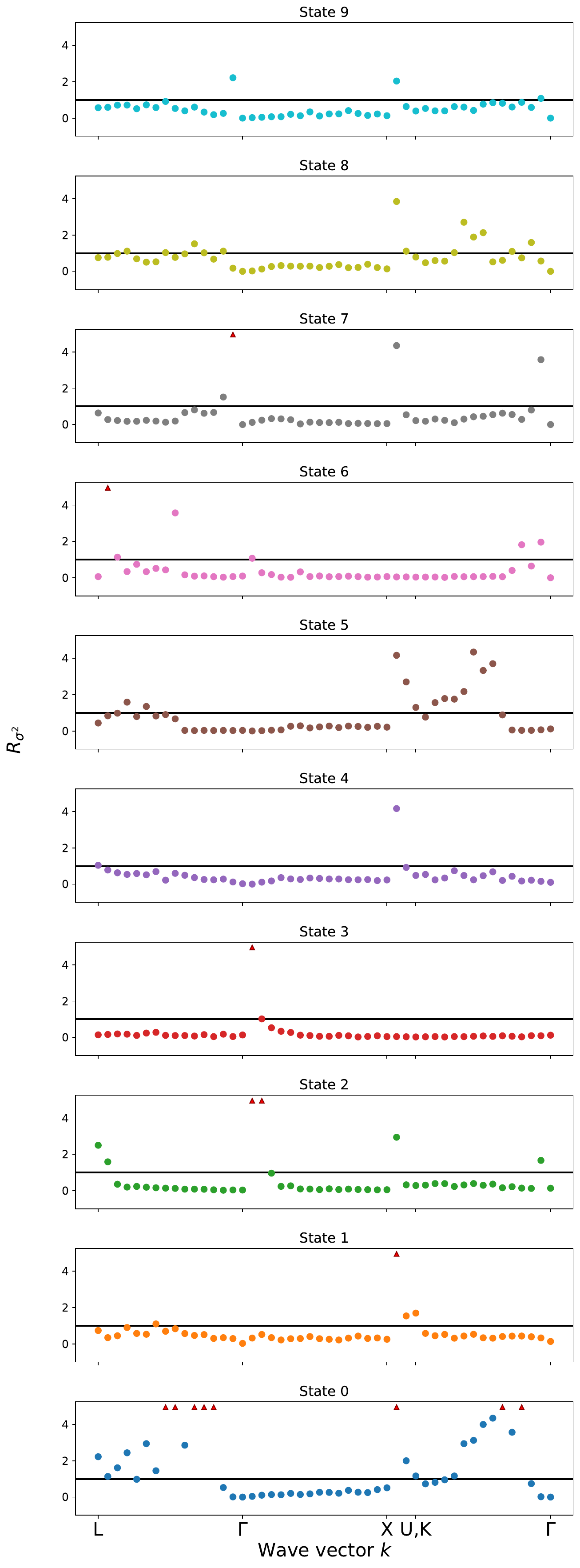} 
    \caption{\textcolor{black}{Variance ratios $R_{\sigma^2}$ for the ten-qubit silicon model as a function of $\boldsymbol{k}$ for each eigenstate. The red triangles denote the outlier ratios where $R_{\sigma^2} > 5$.The horizontal solid black line corresponds to the value of 1, values under this threshold indicate a better performance by the proposed constant-measurement protocol.}}
    \label{fig.R_Si}
\end{figure}

 Overall, the constant protocol appears to estimate the cost function more accurately in most cases than the conventional OWC grouping protocol. To quantify this more precisely, we have computed the variance ratios (\ref{eq.var_ratio}) for each calculated state and displayed the results in Figures \ref{fig.R_CuO2}, \ref{fig.R_bg}, and \ref{fig.R_Si}. For $R_{\sigma^2} > 1$, the conventional protocol is more accurate. $R_{\sigma^2} < 1$, the constant protocol yields better precision. As can be seen from the data,  the ratios $R_{\sigma^2}$ from the four- and ten-qubit models are, in most cases, smaller than 1, indicating better precision for the constant protocol.

\subsection{\textcolor{black}{Extension to Many-Body Fermionic Systems}}

\textcolor{black}{To simulate realistic condensed matter systems, extending the single-particle approximation to the many-body regime is essential. Preparing a non-interacting many-body initial state is a vital precursor for various quantum algorithms. In adiabatic quantum computation, for example, one initialises the system in the ground state of a non-interacting Hamiltonian before slowly evolving it under the full interacting Hamiltonian. Similarly, this non-interacting ground state can serve as an excellent starting point for the Quantum Krylov Diagonalisation \cite{Yoshioka}. Consequently, developing an efficient protocol to prepare non-interacting many-body ground states is of paramount importance. To contextualise these efficiencies, consider a non-interacting part of the multi-orbital Hubbard model in real space, given by the Hamiltonian}

\begin{equation}
    \label{eq:many-body-real}
    \mathcal{\hat{H}} = \sum_{m}\sum_{j\sigma}\varepsilon_{j}\hat{c}^{\dagger}_{m\sigma j}\hat{c}_{m\sigma j} + \sum_{mn}\sum_{jl}\sum_{\sigma}t_{mn}^{jl}\hat{{c}}^{\dagger}_{m\sigma j}\hat{c}_{\sigma l},
\end{equation}

\noindent where $\varepsilon_j$ are on-site energies of orbital $j$ the $t_{mn}^{jl}$ are hopping matrix elements between  orbital on site $j$ on site $m$ and orbital $l$ on site $n$. Representing the full multi-electron system on a quantum computer requires a global register allocation of $2NL$ physical qubits, where $N$ is the number of orbitals in a unit cell and $L$ is the number of unit cells in the crystal. The factor of two accounts for the spin-up and spin-down degrees of freedom. Because kinetic hopping terms couple adjacent spatial sites across the lattice, the real-space register cannot be modularly partitioned. One can exploit spin degeneracy in the non-interacting limit to decouple the spin-up and spin-down sectors, which strategically reduces the required $2NL$ qubits down to $NL$ qubits (or simply $L$ qubits in the single-orbital $N=1$ limit). However, even with this reduction, the optimisation space remains bound to the physical size of the lattice. Simulating a moderately sized crystal to mitigate finite-size errors (e.g., $L = 50$) necessitates executing an iterative Variational Quantum Eigensolver (VQE) loop on a globally entangled $50N$-qubit register. In stark contrast, the momentum-space ($\boldsymbol{k}$-space) formulation completely circumvents these extensive scaling limitations by absorbing the crystal size $L$ classically as a discrete momentum space index. Because the non-interacting multi-orbital Hubbard Hamiltonian, see Eq. (24) in the Supplementary Notes 6, natively block-diagonalises in $\boldsymbol{k}$-space, the global quantum register factorises cleanly into $2L$ independent sub-registers of size $N$, as shown in the Supplementary Note 6. Crucially, this complete factorisation collapses the active qubit requirement during the optimisation phase from an extensive $NL$-qubit problem in real space down to an isolated, localised block of exactly $N$ qubits in momentum space.

The Hamiltonian, Eq. \eqref{eq.ham_qubit}, is a single-particle version of this many-body non-interacting part of the multi-orbital Hubbard model, acting only on $N$ qubits instead of $2NL$. The main idea is that for a non-interacting many-body system, instead of solving one many-body problem on $2NL$ qubits, one can instead solve many one-body problems in just $N$ qubits, see Supplementary Notes 6, where we provide a detailed description of the construction of many-body eigenstates. Crucially, the constant measurement protocol provides an efficient variational quantum algorithm for single-particle energies calculation as well as single-particle eigenstates, from which one can construct the exact eigenstates of non-interacting many-body multi-orbital Hubbard model. Crucially, the circuit depth required to generate these exact many-body eigenstates scales strictly as $O(N)$, determined solely by the number of orbitals within a single unit cell and remains invariant to the global crystal size $L$. Furthermore, the maximum length of any internal Jordan-Wigner $\hat{Z}$-string is tightly bounded by $N-2$, completely eliminating the extensive string penalties associated with large spatial lattices. For realistic materials where the number of orbitals per unit cell is small (typically $N \le 5$), solving for the variational angles that diagonalise an $N \times N$ matrix is computationally trivial and can be performed instantaneously via classical pre-diagonalisation or with the classical variational method. Whether one uses classical routines or independent single-particle quantum optimisation loops, as described in this work, is merely a matter of logistical convenience rather than a computational scaling bottleneck.

On the other hand, there is a fundamental representation trade-off for the simulations on near-term hardware. While the non-interacting background can be optimised effortlessly within a unit cell with $O(N)$ depth in momentum space, the reintroduction of local electron-electron interactions—such as the intra-orbital Coulomb repulsion $U$, inter-orbital repulsion $U'$, and Hund's exchange couplings $J, J'$—alters the scaling landscape dramatically. In a real-space representation, these interaction terms remain strictly local, low-depth density-density operators requiring only local, two-qubit gates. Conversely, transforming these localised interaction terms into the $\boldsymbol{k}$-space basis maps them onto a four-body, all-to-all scattering network. Recognising this duality, our framework offers a highly scalable, hybrid strategy that combines the optimal features of both representations. Rather than enduring the extensive optimisation bottleneck in real space or the dense interaction gates in momentum space, one can efficiently compile the exact non-interacting many-body initial state natively within the localised, parallelised $\boldsymbol{k}$-space registers. Once this state is prepared with $O(N)$ depth, a hardware-level Fermionic Fast Fourier Transform (FFFT) circuit \cite{Babbush} can be executed across the register. The FFFT acts as a global basis transformation that maps the fully antisymmetrised momentum-space eigenstates directly into their real-space counterparts on the hardware.

\clearpage

\section{Conclusion}\label{sec5}
In this work, we introduced a constant measurement protocol for the \textcolor{black}{Orthogonal-Ansatz Variational Quantum Eigensolver} algorithm applied to tight-binding Hamiltonians. By exploiting symmetry properties, we demonstrated that the number of required global measurement settings can be reduced to a constant three, independent of system size. This represents a substantial reduction in measurement overhead compared to conventional approaches that scale linearly with the number of qubits. As benchmarks, we have shown the performance of our approach for \textcolor{black}{three} tight-binding models: a three-qubit model of a two-dimensional CuO$_2$ square lattice with a three-atom basis, a four-qubit model of a two-dimensional bilayer graphene system, \textcolor{black}{and a ten-qubit three-dimensional diamond silicon structure}. \textcolor{black}{We have also demonstrated the stability of the measurement protocol by estimating the Hamiltonian cost function independently of the optimization loop and calculating the standard deviations across the band structure. We found that our constant protocol can estimate the cost function with greater precision than conventional methods, particularly for larger systems (e.g., $N>3$). However, on some occasions, the non-linearity of the product rule in Eq. \eqref{eq.prod_rule} can significantly amplify noise, especially in regimes with low shot counts.}

\textcolor{black}{Finally, we have shown how the protocol can be generalised to construct the exact many-body eigenstates for the non-interacting multi-orbital Hubbard model. By mapping the system to momentum space, the many-body initialisation collapses to independent blocks requiring only $O(N)$ circuit depth and zero scaling overhead with respect to the total crystal size $L$. This state can subsequently be mapped directly into real space via a hardware-level Fermionic Fast Fourier Transform (FFFT) circuit, offering a highly scalable initialisation framework for downstream interacting quantum algorithms while completely bypassing long-range Jordan-Wigner string penalties.}

\backmatter

\bmhead{Data availability}
Data for these experiments is available in Ref. \cite{quantum_vqd}.

\bmhead{Code availability}
Code for these experiments is available in Ref. \cite{quantum_vqd}.

\bmhead{Author Contributions Statement} 
All authors contributed to writing and reviewing the manuscript. The measurement protocol was primarily developed by M.K., with software development support from L.K. and additional contributions from I.A.M., M.P., and M.F.

\bmhead{Competing Interests Statement}
The authors declare that they have no competing interests or other interests that might be perceived to influence the results and/or discussion reported in this paper. 

\bmhead{Funding declaration}
The authors acknowledge financial support from the Czech Academy of Sciences ({\it Praemium Academiae} and the Strategy AV21, in particular the program "AI: Artificial Intelligence for Science and Society") and the Ministry of Education, Youth and Sports of the Czech Republic (project No. LUC25028  within the INTER-EXCELLENCE II program, a subprogram INTER-COST - LUC25). M.P. and I.A.M. acknowledge the support by the VEGA project No. 2/0055/23 as well as from the Research and Innovation Authority projects 09I03-03-V04-00425 and 09I03-03-V04-00685.

\bmhead{Acknowledgements}
 Computational resources were made available by the Ministry of Education, Youth and Sports of the Czech Republic under the Project e-INFRA CZ (ID:90254) at the IT4Innovations National Supercomputing Center, the MetaCentrum and CERIT-SC. Access to CESNET storage facilities provided by the project e-INFRA CZ under the program "Projects of Large Research, Development and Innovations Infrastructures" (LM2023054). This research was supported by the Quantum Innovation Center (QIC) project under the Consortium Agreement for the Use of Quantum Technology Services No. UO/09/007/2025 as well as the support by the Quantum Innovation Center QIC-Czech, is appreciated.


\bibliography{sn-bibliography}

@article{Schuch,
    author = "Schuch, N. and Verstraete, F.",
    title = "Computational complexity of interacting electrons and fundamental limitations of density functional theory." ,
    journal = "Nature Phys",
    volume = "5",
    pages = "732–735" ,
    year = "2009",
    doi = "10.1038/nphys1370",
}

@article{Yoshioka,
    author = "Yoshioka, N. and Amico, M. and Kirby, W. et al.",
    title = "Krylov diagonalization of large many-body Hamiltonians on a quantum processor." ,
    journal = "Nat Commun",
    volume = "16",
    pages = "5014" ,
    year = "2025",
    doi = "10.1038/s41467-025-59716-z",
    url = {https://doi.org/10.1038/s41467-025-59716-z}
}

@article{Babbush,
  title = {Low-Depth Quantum Simulation of Materials},
  author = {Babbush, Ryan and Wiebe, Nathan and McClean, Jarrod and McClain, James and Neven, Hartmut and Chan, Garnet Kin-Lic},
  journal = {Phys. Rev. X},
  volume = {8},
  issue = {1},
  pages = {011044},
  numpages = {40},
  year = {2018},
  month = {Mar},
  publisher = {American Physical Society},
  doi = {10.1103/PhysRevX.8.011044},
  url = {https://link.aps.org/doi/10.1103/PhysRevX.8.011044}
}

@article{Nakanishi,
  title = {Subspace-search variational quantum eigensolver for excited states},
  author = {Nakanishi, Ken M. and Mitarai, Kosuke and Fujii, Keisuke},
  journal = {Phys. Rev. Res.},
  volume = {1},
  issue = {3},
  pages = {033062},
  numpages = {7},
  year = {2019},
  month = {Oct},
  publisher = {American Physical Society},
  doi = {10.1103/PhysRevResearch.1.033062},
  url = {https://link.aps.org/doi/10.1103/PhysRevResearch.1.033062}
}

@article{Cohen,
    author = "Cohen, Aron J. and Mori-Sánchez, Paula and Yang, Weitao",
    title = "Challenges for Density Functional Theory." ,
    journal = "Chem. Rev.",
    volume = "112",
    pages = "289-320" ,
    year = "2012",
    doi = "10.1021/cr200107z",
}

@article{Whitfield,
  author  = {Whitfield, James Daniel and Love, Peter John and Aspuru-Guzik, Al{\\'a}n},
  title   = {Computational complexity in electronic structure},
  journal = {Phys. Chem. Chem. Phys.},
  volume  = {15},
  pages   = {397--411},
  year    = {2013},
  doi     = {10.1039/C2CP42695A}
}

@article{SK,
  author  = {Slater, J. C. and Koster, G. F.},
  title   = {Simplified LCAO Method for the Periodic Potential Problem},
  journal = {Phys. Rev.},
  volume  = {94},
  pages   = {1498--1524},
  year    = {1954},
  doi     = {10.1103/PhysRev.94.1498}
}

@article{Chadi,
  author  = {Chadi, D. J. and Cohen, M. L.},
  title   = {Tight-Binding Calculations of the Valence Bands of Diamond and Zincblende Crystals},
  journal = {Phys. Status Solidi B},
  volume  = {68},
  pages   = {405--419},
  year    = {1975},
  doi     = {10.1002/pssb.2220680140}
}

@book{Harrison,
  author    = {Harrison, Walter A.},
  title     = {Electronic Structure and the Properties of Solids: The Physics of the Chemical Bond},
  publisher = {Dover Publications},
  year      = {1989}
}

@article{Peruzzo,
  author  = {Peruzzo, Alberto and McClean, Jarrod and Shadbolt, Peter and Yung, Man-Hong and Zhou, Xiao-Qi and Love, Peter J. and Aspuru-Guzik, Alan and O'Brien, Jeremy L.},
  title   = {A variational eigenvalue solver on a photonic quantum processor},
  journal = {Nat. Commun.},
  volume  = {5},
  pages   = {4213},
  year    = {2014},
  doi     = {10.1038/ncomms5213}
}

@article{Higgott,
  author  = {Higgott, Oscar and Wang, Daochen and Brierley, Stephen},
  title   = {Variational Quantum Computation of Excited States},
  journal = {Quantum},
  volume  = {3},
  pages   = {156},
  year    = {2019},
  doi     = {10.22331/q-2019-07-01-156}
}

@article{Bittel,
  author  = {Bittel, Lennart and Kliesch, Martin},
  title   = {Training Variational Quantum Algorithms Is NP-Hard},
  journal = {Phys. Rev. Lett.},
  volume  = {127},
  pages   = {120502},
  year    = {2021},
  doi     = {10.1103/PhysRevLett.127.120502}
}

@article{Bonet,
  author  = {Bonet-Monroig, Xavier and Wang, Hao and Vermetten, Diederick and Senjean, Bruno and Moussa, Charles and B{\"a}ck, Thomas and Dunjko, Vedran and O'Brien, Thomas E.},
  title   = {Performance comparison of optimization methods on variational quantum algorithms},
  journal = {Phys. Rev. A},
  volume  = {107},
  pages   = {032407},
  year    = {2023},
  doi     = {10.1103/PhysRevA.107.032407}
}

@article{Kandala,
  author  = {Kandala, Abhinav and Mezzacapo, Antonio and Temme, Kristan and Takita, Maika and Brink, Markus and Chow, Jerry M. and Gambetta, Jay M.},
  title   = {Hardware-efficient variational quantum eigensolver for small molecules and quantum magnets},
  journal = {Nature},
  volume  = {549},
  pages   = {242--246},
  year    = {2017},
  doi     = {10.1038/nature23879}
}

@article{McClean,
  author  = {McClean, Jarrod R. and Romero, Jonathan and Babbush, Ryan and Aspuru-Guzik, Al{\'a}n},
  title   = {The theory of variational hybrid quantum-classical algorithms},
  journal = {New J. Phys.},
  volume  = {18},
  pages   = {023023},
  year    = {2016},
  doi     = {10.1088/1367-2630/18/2/023023}
}

@article{Sherbert1,
  author  = {Sherbert, Kyle and Cerasoli, Frank and Buongiorno Nardelli, Marco},
  title   = {A systematic variational approach to band theory in a quantum computer},
  journal = {RSC Adv.},
  volume  = {11},
  pages   = {39438--39449},
  year    = {2021},
  doi     = {10.1039/D1RA07451B}
}

@article{Sherbert2,
  author  = {Sherbert, Kyle and Jayaraj, A. and Buongiorno Nardelli, Marco},
  title   = {Quantum algorithm for electronic band structures with local tight-binding orbitals},
  journal = {Sci. Rep.},
  volume  = {12},
  pages   = {9867},
  year    = {2022},
  doi     = {10.1038/s41598-022-13627-x}
}

@article{Goringe,
  author  = {Goringe, C. M. and Bowler, D. R. and Hern{\'a}ndez, E.},
  title   = {Tight-binding modelling of materials},
  journal = {Rep. Prog. Phys.},
  volume  = {60},
  pages   = {1447},
  year    = {1997},
  doi     = {10.1088/0034-4885/60/12/001}
}

@article{Wannier,
  author  = {Marzari, Nicola and Mostofi, Arash A. and Yates, Jonathan R. and Souza, Ivo and Vanderbilt, David},
  title   = {Maximally localized Wannier functions: Theory and applications},
  journal = {Rev. Mod. Phys.},
  volume  = {84},
  pages   = {1419--1475},
  year    = {2012},
  doi     = {10.1103/RevModPhys.84.1419}
}

@article{Nature-Phys-2023-Zhu,
  author  = {Zhu, D. and Kahanamoku-Meyer, G.D. and Lewis, L. {\it et al.} },
  title   = {Interactive cryptographic proofs of quantumness using mid-circuit measurements},
  journal = {Nat. Phys. },
  volume  = {19},
  pages   = {1725--1731},
  year    = {2023},
  doi     = {10.1038/s41567-023-02162-9}
}

@article{Duriska,
  author  = {Ďuriška, Michal and Miháliková, Ivana and Friák, Martin},
  title   = {Quantum computing of the electronic structure of crystals by the Variational Quantum Deflation algorithm},
  journal = {Phys. Scr.},
  volume  = {100},
  pages   = {045105},
  year    = {2025},
  doi     = {10.1088/1402-4896/adbb29}
}

@article{Kuzmenko,
  author  = {Kuzmenko, A. B. and Crassee, I. and van der Marel, D. and Blake, P. and Novoselov, K. S.},
  title   = {Determination of the gate-tunable band gap and tight-binding parameters in bilayer graphene using infrared spectroscopy},
  journal = {Phys. Rev. B},
  volume  = {80},
  pages   = {165406},
  year    = {2009},
  doi     = {10.1103/PhysRevB.80.165406}
}

@Inbook{Fulde1995,
author="Fulde, P.",
title="Superconductivity and the High-Tc Materials",
bookTitle="Electron Correlations in Molecules and Solids",
year="1995",
publisher="Springer Berlin Heidelberg",
address="Berlin, Heidelberg",
pages="377--422",
isbn="978-3-642-57809-0",
doi="10.1007/978-3-642-57809-0_14",
url="https://doi.org/10.1007/978-3-642-57809-0_14"
}

@phdthesis{Ragonneau,
  author = {Ragonneau, T. M.},
  title  = {Model-Based Derivative-Free Optimization Methods and Software},
  school = {Department of Applied Mathematics, The Hong Kong Polytechnic University},
  year   = {2022},
  url    = {https://theses.lib.polyu.edu.hk/handle/200/12294}
}

@article{SciPy,
  author  = {Virtanen, Pauli and Gommers, Ralf and Oliphant, Travis E. and Haberland, Matt and Reddy, Tyler and Cournapeau, David and Burovski, Evgeni and Peterson, Pearu and Weckesser, Warren and Bright, Jonathan and {et al.}},
  title   = {SciPy 1.0: fundamental algorithms for scientific computing in Python},
  journal = {Nat. Methods},
  volume  = {17},
  pages   = {261--272},
  year    = {2020},
  doi     = {10.1038/s41592-019-0686-2}
}

@article{Qiskit2024,
  author  = {Javadi-Abhari, Ali and Treinish, Matthew and Krsulich, Kevin and Wood, Christopher J. and Lishman, Jake and Gacon, Julien and Martiel, Simon and Nation, Paul and Bishop, Lev S. and Cross, Andrew W. and Johnson, Blake R. and Gambetta, Jay M.},
  title   = {Quantum computing with Qiskit},
  journal = {arXiv preprint arXiv:2405.08810},
  year    = {2024},
  url     = {https://arxiv.org/abs/2405.08810}
}

@article{McCann,
doi = {10.1088/0034-4885/76/5/056503},
url = {https://dx.doi.org/10.1088/0034-4885/76/5/056503},
year = {2013},
month = {apr},
publisher = {IOP Publishing},
volume = {76},
number = {5},
pages = {056503},
author = {McCann, Edward and Koshino, Mikito},
title = {The electronic properties of bilayer graphene},
journal = {Reports on Progress in Physics}
}

@misc{quantum_vqd,
  author = {Krejci, M},
  title = {quantum\_vqd},
  year = {2025},
  publisher = {GitHub},
  journal = {GitHub repository},
  url = {https://github.com/codebykrejci/quantum_vqd}
}

@article{cj89-4h5t,
  title = {Readout Error Mitigation for Mid-Circuit Measurements and Feedforward},
  author = {Koh, Jin Ming and Koh, Dax Enshan and Thompson, Jayne},
  journal = {PRX Quantum},
  volume = {7},
  issue = {1},
  pages = {010317},
  numpages = {40},
  year = {2026},
  month = {Jan},
  publisher = {American Physical Society},
  doi = {10.1103/cj89-4h5t},
  url = {https://link.aps.org/doi/10.1103/cj89-4h5t}
}

@article{Kaldenbach_2025,
doi = {10.1088/2058-9565/ad802b},
url = {https://doi.org/10.1088/2058-9565/ad802b},
year = {2024},
month = {oct},
publisher = {IOP Publishing},
volume = {10},
number = {1},
pages = {015010},
author = {Kaldenbach, Thierry N and Heller, Matthias},
title = {Mapping quantum circuits to shallow-depth measurement patterns based on graph states},
journal = {Quantum Science and Technology}
}

@article{Altuntas_2025,
doi = {10.1088/2058-9565/ade335},
url = {https://doi.org/10.1088/2058-9565/ade335},
year = {2025},
month = {jun},
publisher = {IOP Publishing},
volume = {10},
number = {3},
pages = {035045},
author = {Altuntaş, E and Lena, R G and Flannigan, S and Daley, A J and Spielman, I B},
title = {Dynamical structure factor from weak measurements},
journal = {Quantum Science and Technology}
}

@article{Romanova_2026,
doi = {10.1088/2058-9565/ae3b6f},
url = {https://doi.org/10.1088/2058-9565/ae3b6f},
year = {2026},
month = {feb},
publisher = {IOP Publishing},
volume = {11},
number = {1},
pages = {015054},
author = {Romanova, Alena and Dür, Wolfgang},
title = {Measurement-based quantum computing with qudit stabilizer states},
journal = {Quantum Science and Technology}
}

@article{Ballesteros-Ferraz_2024,
doi = {10.1088/2058-9565/ad420b},
url = {https://doi.org/10.1088/2058-9565/ad420b},
year = {2024},
month = {may},
publisher = {IOP Publishing},
volume = {9},
number = {3},
pages = {035029},
author = {Ballesteros Ferraz, Lorena and Martin, John and Caudano, Yves},
title = {On the relevance of weak measurements in dissipative quantum systems},
journal = {Quantum Science and Technology}
}

@article{Caprotti_2024,
doi = {10.1088/2058-9565/ad4584},
url = {https://doi.org/10.1088/2058-9565/ad4584},
year = {2024},
month = {may},
publisher = {IOP Publishing},
volume = {9},
number = {3},
pages = {035032},
author = {Caprotti, A and Barbiero, M and Tarallo, M G and Genoni, M G and Bertaina, G},
title = {Analysis of spin-squeezing generation in cavity-coupled atomic ensembles with continuous measurements},
journal = {Quantum Science and Technology}
}

@article{Burns_2025,
doi = {10.1088/2058-9565/ad9d74},
url = {https://doi.org/10.1088/2058-9565/ad9d74},
year = {2024},
month = {dec},
publisher = {IOP Publishing},
volume = {10},
number = {1},
pages = {015054},
author = {Burns, Matthew X and Liu, Chenxu and Stein, Samuel and Peng, Bo and Kowalski, Karol and Li, Ang},
title = {GALIC: hybrid multi-qubitwise pauli grouping for quantum computing measurement},
journal = {Quantum Science and Technology}
}

@article{Ge_2025,
doi = {10.1088/2058-9565/ad97d8},
url = {https://doi.org/10.1088/2058-9565/ad97d8},
year = {2024},
month = {dec},
publisher = {IOP Publishing},
volume = {10},
number = {1},
pages = {015046},
author = {Ge, Chengfang and Zhou, Lai and Lin, Jinping and Yin, Hua-Lei and Zeng, Qiang and Yuan, Zhiliang},
title = {Post-measurement pairing quantum key distribution with local optical frequency standard},
journal = {Quantum Science and Technology}
}

@article{Nie_2024,
doi = {10.1088/2058-9565/ad34f4},
url = {https://doi.org/10.1088/2058-9565/ad34f4},
year = {2024},
month = {apr},
publisher = {IOP Publishing},
volume = {9},
number = {2},
pages = {025024},
author = {Nie, You-Qi and Zhou, Hongyi and Bai, Bing and Xu, Qi and Ma, Xiongfeng and Zhang, Jun and Pan, Jian-Wei},
title = {Measurement-device-independent quantum random number generation over 23 Mbps with imperfect single-photon sources},
journal = {Quantum Science and Technology}
}

@article{Lee_2024,
doi = {10.1088/2058-9565/ad6d87},
url = {https://doi.org/10.1088/2058-9565/ad6d87},
year = {2024},
month = {aug},
publisher = {IOP Publishing},
volume = {9},
number = {4},
pages = {045028},
author = {Lee, Jinil and Song, Wooyeong and Lee, Donghwa and Kim, Yosep and Lee, Seung-Woo and Lim, Hyang-Tag and Jung, Hojoong and Han, Sang-Wook and Kim, Yong-Su},
title = {Photonic variational quantum eigensolver using entanglement measurements},
journal = {Quantum Science and Technology}
}

@article{plesch2011quantum,
  title={Quantum-state preparation with universal gate decompositions},
  author={Plesch, Martin and Brukner, {\v{C}}aslav},
  journal={Physical Review A—Atomic, Molecular, and Optical Physics},
  volume={83},
  number={3},
  pages={032302},
  year={2011},
  publisher={APS}
}


\end{document}